\documentclass[iop, tighten, numberedappendix,letter]{emulateapj}

\usepackage{graphicx}
\usepackage{latexsym,amssymb}
\usepackage{amsmath}
\usepackage{natbib}

\newcommand{\msig}{M$_\bullet$-$\sigma$}
\newcommand{\kms}{km\,s$^{-1}$}

\newcommand{\Msun}{M$_\odot$}
\newcommand{\Mbh}{$M_\bullet$}

\newcommand{\brg}{Br$\gamma$}
\renewcommand{\deg}{\ensuremath{^{\circ}}}

\slugcomment{Draft version \today}

\shorttitle{The low mass MBH of NGC4395} 

\shortauthors{den Brok, et al.}

\begin{document}

\title{Measuring the mass of the central black hole in the bulgeless galaxy NGC 4395 from gas dynamical modeling}

\author{Mark~den~Brok\altaffilmark{1}}
\author{Anil~C.~Seth\altaffilmark{1}}
\author{Aaron~J.~Barth\altaffilmark{2}}
\author{Daniel~J.~Carson\altaffilmark{2}}
\author{Nadine~Neumayer\altaffilmark{3}}
\author{Michele Cappellari\altaffilmark{4}}
\author{Victor~P.~Debattista\altaffilmark{5}}
\author{Luis~C.~Ho\altaffilmark{6,7}}
\author{Carol~E.~Hood\altaffilmark{8}}
\author{Richard~M.~McDermid\altaffilmark{9,10}}

\altaffiltext{1}{Department of Physics and Astronomy, University of Utah, Salt Lake City, Utah 84112, USA: denbrok@physics.utah.edu}
\altaffiltext{2}{Department of Physics and Astronomy, 4129 Frederick Reines Hall, University of California, Irvine, CA 92697-4575, USA}
\altaffiltext{3}{Max-Planck Institute for Astronomy, K\"onigstuhl 17, 69117 Heidelberg, Germany}
\altaffiltext{4}{Sub-department of Astrophysics, Department of Physics, University of Oxford, Denys Wilkinson Building, Keble Road, Oxford, OX1 3RH, UK}
\altaffiltext{5}{Jeremiah Horrocks Institute, University of Central Lancashire, Preston, Lancashire, PR1 2HE, UK}
\altaffiltext{6}{Kavli Institute for Astronomy and Astrophysics, Peking University, Beijing 100871, China}
\altaffiltext{7}{Department of Astronomy, School of Physics, Peking University, Beijing 100871, China}
\altaffiltext{8}{Department of Physics, California State University, San Bernardino, 5500 University Parkway, San Bernardino CA 92407-2318, USA}
\altaffiltext{9}{Department of Physics and Astronomy, Macquarie University, Sydney NSW 2109, Australia}
\altaffiltext{10}{Australian Gemini Office, Australian Astronomical Observatory, PO Box 915, Sydney NSW 1670, Australia}

\begin{abstract}
NGC 4395 is a bulgeless spiral galaxy, harboring one of the nearest known type 1 Seyfert nuclei. Although there is no consensus on the mass of its central engine, several estimates suggest it to be one of the lightest massive black holes (MBHs) known. We present the first direct dynamical measurement of the mass of this MBH from a combination of two-dimensional gas kinematic data, obtained with the adaptive optics assisted near infrared integral field spectrograph Gemini/NIFS, and high-resolution multiband photometric data from {\it Hubble Space Telescope}'s Wide Field Camera 3 ({\it HST}/WFC3). We use the photometric data to model the shape and stellar mass-to-light ratio (M/L) of the nuclear star cluster. From the Gemini/NIFS observations, we derive the kinematics of warm molecular hydrogen gas as traced by emission through the H$_2$ 1--0 S(1) transition. These kinematics show a clear rotational signal, with a position angle orthogonal to NGC 4395's radio jet. Our best fitting tilted ring models of the kinematics of the molecular hydrogen gas contain a black hole with mass $M=4_{-3}^{+8}\times 10^5$  \Msun (3$\sigma$ uncertainties) embedded in a nuclear star cluster of mass $M=2 \times 10^6$  \Msun. Our black hole mass measurement is in excellent agreement with the reverberation mapping mass estimate of Peterson et al. (2005), but shows some tension with other mass measurement methods based on accretion signals.
\end{abstract}


\section{Introduction}
\label{sec:intro}

It is suspected that almost all high-mass galaxies harbor a supermassive black hole at their center \citep[e.g.][]{KorRic95,KorHo13}. There are two popular scenarios to form such massive black holes. In the first scenario, black holes are created as stellar remnants, most likely from heavy Population III stars \citep[e.g.][]{BroLar04, Glo05, Bro13}. The second option is a scenario in which primordial black hole seeds form from direct collapse of gas clouds leading to $10^4$--$10^5$ \Msun\ black holes \citep[e.g.][]{HaeRee93,KouBulDek04,BegVolRee06}. Both scenarios would require a period of rapid growth for the seed black hole in order to produce the massive quasars found at high redshifts \citep{FanNarLup01,JiaFanBia09,WilDelRey09,VenFinSut13}. 

Due to subsequent growth of these seeds, both black hole formation scenarios predict high present day occupation fraction and the same black hole mass functions in high-mass galaxies. 
However, the direct collapse scenario predicts a much lower occupation fraction in low-mass galaxies than the remnant scenario \citep{VolLodNat08,Gre12}. A census of low-mass supermassive black holes in low-mass galaxies can thus provide important constraints on the formation mechanisms of supermassive black holes in the early Universe.
For high-mass galaxies, the mass of the MBH appears to correlate tightly with the properties of the bulge. Specifically, there is a correlation with the luminosity and mass of the bulge \citep{KorRic95,MagTreRic98}, but the strongest correlation was found to be between black hole mass and velocity dispersion of the stars in the bulge, the \msig\ relation \citep{FerMer00, GebBenBow00}. The tightness of this relation implies some physical connection between the evolution of bulges and MBHs. Combined with the lack of a massive black hole in M33, this led to the suggestion that black holes may be absent from pure disk systems \citep{MerFerJos01, GebLauKor01}. 
The presence of a black hole in NGC4395 defies this expectation.  The galaxy is classified as a Sd galaxy \citep{SanTam81}, thus with essentially no bulge (see also Appendix \ref{apx:decomposition}), but, from its broad nuclear emission lines, has been observed to contain a Seyfert 1 nucleus \citep{FilSar89}. A measurement of the mass of the central black hole in NGC 4395 helps our understanding of the co-evolution of galaxies and black holes in multiple ways - by adding to the statistics of MBHs in low-mass galaxies and thus understanding the M-$\sigma$ relation and its scatter at low masses,  by showing that MBHs can be present even in the complete absence of a bulge, and by exploring the demographics of MBHs in spiral galaxies.

The mass of the black hole in NGC 4395 has been estimated by a variety of methods. Using reverberation mapping, \citet{PetBenDes05} estimated the mass to be \Mbh $ = 3.6 \times 10^5$ \Msun. However, more recently, \citet{EdrRafChe12} estimated the black hole mass using reverberation mapping to find \Mbh $ = (4.9 \pm 2.6) \times 10^4$ \Msun; the different mass estimates are almost entirely due to modeling assumptions. The conversion of a time delay to a black hole mass is based on  a virial factor dependent on the broad-line region (BLR) geometry.  This virial factor is often calibrated by assuming that reverberation mapping black holes masses follow the same \msig\ relation as those in inactive galaxies for which BH masses can be measured from stellar dynamics \citep[e.g.][]{OnkFerMer04}. 
The estimates for NGC 4395's black hole mass therefore rely on two assumptions: (1) that the \msig\ relation for quiescent, mostly early-type galaxies is still valid for the spiral galaxies typically probed by reverberation mapping, despite evidence of the contrary \citep{GrePenKim10}, and (2) that the BLR geometry remains similar at the lower mass inferred in NGC 4395.  Evidence for a different mode of accretion has been suggested by NGC 4395's X-ray luminosity and variability (\citealt{MorFilHo99,MorEraLei05}; but see \citealt{NarRis11}).

Obtaining a mass measurement of the MBH in NGC 4395 in an independent and more direct way, such as dynamical modeling, is therefore important. However, the masses of BHs in low-mass galaxies are much harder to measure than in high-mass galaxies, partly because of the lower surface brightness of the host galaxy and because the sphere of influence, in which the gravitional potential is dominated by the mass of the black hole, is smaller, but often also because of radially varying stellar mass-to-light ratios (M/L) due to recent star formation or the presence of a nuclear star cluster \citep[NSC,][]{WalBokCha06,RosvanBok06,SetDalHod06,CarBarSet15}.

Several authors have inferred black hole masses from modeling the nuclear gas kinematics in nearby massive galaxies \citep[][]{MacMarAxo97,vanvan98,VervanCar00,NeuCapReu07,BarSarRix01,SarRixShi01,CapVervan02,ShaCapdeZ06}. Here we present a dynamical measurement of the mass of the MBH and nuclear star cluster in NGC 4395, using high-spatial resolution adaptive optics assisted observations of the molecular gas kinematics from Gemini/NIFS. This is the lowest mass MBH and, besides M60-UCD1 \citep{SetvanMie14}, the lowest mass galaxy for which the black hole mass has been measured by dynamical modeling. 

The structure of this paper is as follows. In Section \ref{sec:data}, we describe the photometric data, which we use to determine the stellar mass density (Sec. \ref{sec:ssp_modeling}), and the spectroscopic data, from which we determine the combined dark and luminous mass density by modeling the gas kinematics (Sec. \ref{sec:gas_modeling}). Throughout this paper, we assume that the distance to NGC 4395 is 4.4 Mpc, which is in between the Cepheid distance \citep[depending on the choice of the period-luminosity relation either 4.0 or 4.3 Mpc,][]{ThiHoeSah04} and tip of the red giant branch (TRGB) distances \citep[4.6 and 4.76 Mpc,][]{KarShaDol03,JacRizTul09}.

\section{Data}
\label{sec:data}

\begin{deluxetable*}{ccccc}
\tablecolumns{4} 
\tablecaption{Summary of HST/WFC3 observations}
\tablehead{     & \colhead{Expose time} & \colhead{Zeropoint} & \colhead{$A_{\lambda}^1$} \\ 
\colhead{Band} & \colhead{[seconds]} & \colhead{[mag]} & \colhead{[mag]}  } 
\startdata
F275W & $8\times150$   & 22.6322 & 0.094\\
F336W & $4\times254$    & 23.4836 & 0.077 \\
F438W & $4\times148$  & 24.9738 & 0.062\\
F547M & $4\times120$   & 24.7477 & 0.050\\
F814W & $4\times100$  & 24.6803 & 0.026
\enddata
\tablerefs{ (1) The extinction values $A_\lambda$ were obtained from \citet{SchFin11}}\label{tab:hstobs}

\end{deluxetable*}
\subsection{Imaging data}
The center of NGC 4395 was imaged with HST/WFC3 in 7 filter bands, ranging from the ultraviolet (UV) to near-infrared (NIR).  Instead of a broad {\it V} band, we used the medium-wide F547M filter, to exclude most bright emission lines. The observations are summarized in Table~\ref{tab:hstobs}. Each filter band consisted of at least 4 exposures with different dither positions. These data are part of a larger survey to infer the formation mechanisms of the nearest nuclear star clusters and have already been presented in \citet{CarBarSet15}, however, to obtain optimal spatial resolution and better noise characteristics, we use a slightly different procedure to reduce the data.

 We drizzled the flat-fielded optical/UV images using the {\it drizzlepac} package in {\sc pyraf}. We found good results in terms of resolution and noise by drizzling the optical images with a Lanczos3 kernel (with {\it pixfrac} set to 1 pixel). The World Coordinate System of the images was aligned to that of the F438W band using the {\it tweakreg} package, after which the aligned flat-fielded images were re-drizzled. The spatial alignment between the optical/UV bands is better than 0.1 pixel.

Although we also imaged NGC 4395 in the NIR with HST/WFC3, we did not use these data for this work, since they have slightly lower resolution, whereas the separation of the AGN contribution to the light of the NSC requires superior resolution.

In addition to the WFC3 data, we also use archival HST/ACS F606W and F814W wide field camera (WFC) data. The F606W filter is important for identifying line emission, since it contains H$\alpha$. The F814W ACS data is necessary to make a F606W-F814W colormap. We drizzled the F606W and F814W data on the same grid as the WFC3 data, so that we were able to use these data to mask emission line regions in the WFC3 images.

\subsubsection{Point spread function}
Point spread functions (PSFs) were generated using {\it TinyTim} \citep{KriHooSto11}. This program generates raw PSFs, with the same geometric distortion as the flatfielded HST/WFC3 images. Since we fit our \textsc{Galfit} models on the distortion-corrected drizzled images, we have to drizzle the PSF in the same way as the flatfielded data. For each band, we created therefore four intermediate PSFs with {\it TinyTim}, with the same pixel offsets as the dither pattern of the actual observations. These intermediate PSFs were injected in the flat-fielded science frames, which we filled with zeroes first. We drizzled these frames with the same parameters as the other imaging data, making sure that no sky subtraction was performed during the drizzling process. We found that, after drizzling, the final PSFs were not centered on a pixel center, and we therefore shifted the four offsets that were used as input for {\it TinyTim} until the drizzled PSFs were centered on a pixel center.

\subsection{Spectroscopic data}

NGC 4395 was observed in the {\it K} band with Gemini/NIFS, a near-infrared image-slicing integral field spectrograph on Gemini-North \citep{McGHarCon03} with ALTAIR laser guide star adaptive optics.  A nearby star was used to tune the slow focus for these observations.  Data was taken on three separate nights in 2010: March 28th, May 5th, and May 22nd.  Exposures were taken in a object-sky-object sequence and a total of 12$\times$600s on source exposures were taken.  A telluric calibrator with spectral type of A0V was taken on each night and used to remove telluric absorption. 

\citet{FilHo03} tried to measure the velocity dispersion of the NSC to constrain the black hole mass, but could only obtain an upper limit on the dispersion of 30 \kms. To get a better measurement of the dispersion, we observed the NGC 4395 nucleus with NIRSPEC at the Keck II telescope. Unfortunately, the CO band heads, which are typical of evolved stellar populations, were absent, probably because of the continuum emission of the AGN, so that we were unable to obtain a dispersion measurement from them. Since we do not use the NIRSPEC data for our analysis, we present the details of these data and the reduction procedures in Appendix \ref{sec:nirspec}, together with the details of our attempt to measure the dispersion from the CO bandheads in the NIFS data.

\subsubsection{NIFS data reduction and PSF determination}
The data was reduced as described by \citet{SetCapNeu10}; initial reduction was done using tools from the IRAF Gemini package that were modified to enable propogation of the variance spectrum. The cubes were rebinned to a pixel size of $0\farcs05\times0\farcs05$ using our own IDL version of NIFCUBE to make final data cubes for each night.  The 12 data cubes were then combined after shifting the May data to match the March 28th data's barycentric velocity.  The line spread function (LSF) of the data was determined by reducing the sky cubes and combining them in the same fashion as the science data; the sky lines in each spatial pixel (spaxel) were then fit to Gaussians to obtain the variation in LSF across the chip; the full width half maximum (FWHM) of the LSF ranged from 4.0 \AA~to 5.0 \AA~across the chip with a median of 4.5 \AA, a typical spectral resolving power of $\lambda/\Delta \lambda \sim 4800$ at the wavelength of the H$_2$ 1-0 S(1) line used in our analysis.  

The intensity, velocity and velocity dispersion of the H$_2$ 1-0 S(1), H$_2$ 1-0 Q(1) and \brg\ lines were determined by fitting each line with a Gaussian. We found that close to the center of the cluster, the \brg\ line did not have a Gaussian shape, but could be fitted by two Gaussians: one with a narrow component ($\sigma \approx 40$ \kms) and one with a broad component ($\sigma \approx 300$ \kms). The width of the broad component suggests the emission originates from the BLR.

\begin{figure*}[t!]
\begin{center}
 \includegraphics[width=1.0\textwidth,clip=]{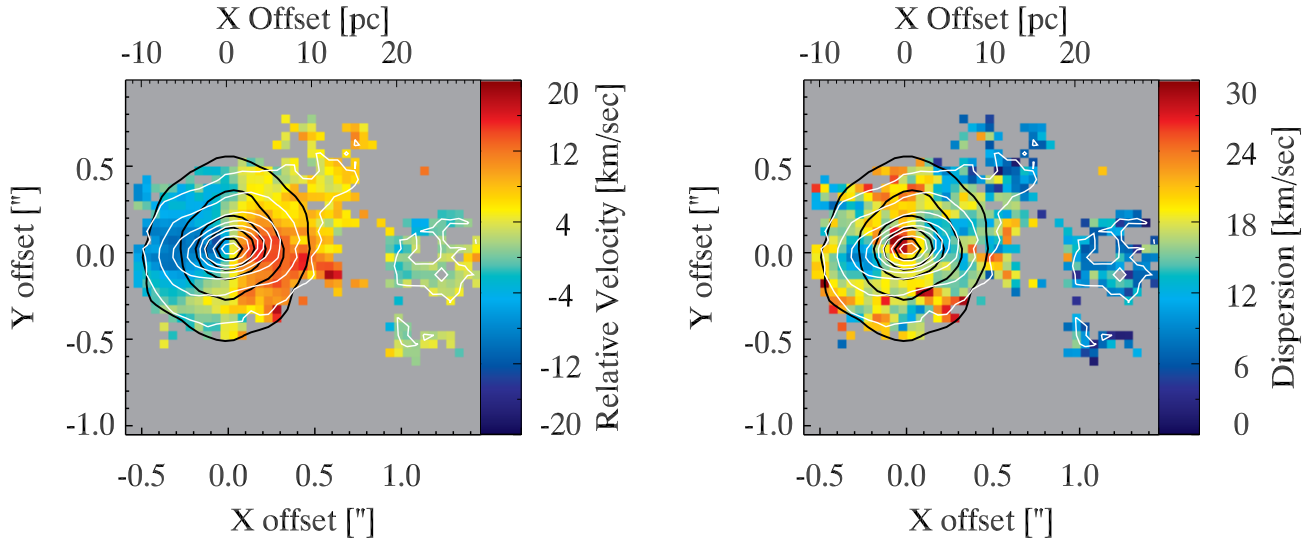}
 \caption{The H$_2$ 1-0 S(1) velocity and velocity dispersion fields. North is up and East is left. White contours represent the intensity of the H$_2$ emission and black contours represent the total intensity in the $K_S$ band.}\label{fig:h2velfield}
\end{center}
\end{figure*}

We used a Bayesian criterion to tell us where the data warranted a two-component fit to the \brg\ line. At these positions, the \brg\ line was fitted by a double Gaussian. To constrain the central potential of NGC 4395, we use the H$_2$ 1-0 S(1), which has the highest S/N of the two H$_2$ lines.
The H$_2$ 1-0 S(1) kinematic maps are shown in Fig. \ref{fig:h2velfield}. Although gas velocity dispersions can only be accurately determined for dispersions higher than $\sim$15 \kms, we note that our results are not sensitive to the dispersion of the disk (see \S5).  

To model the PSF for the kinematics, we use the continuum emission from the AGN. This is justified, since we know from the absence of the CO band heads that the continuum emission is made up almost completely of non-stellar emission (see Appendix \ref{sec:nirspec} for more details). We have checked the width of the continuum against that of the flux from the broad line \brg\ emission and a star in the field. The FWHMs of these other components agree with the value of the continuum emission.

We note that, as in \citet{SetCapNeu10}, the PSF can be modeled as a combination of a Gaussian (with FWHM=0\farcs16), describing the instrumental PSF, and a Moffat profile, describing the seeing: 
$\Sigma(r) = \Sigma_0/[(1+(R/R_d)^2)^{4.765}]$. We use $R_d = 0\farcs95$. The inner core of the PSF is slightly broader than the one used by \citet{SetCapNeu10}, which is probably due to the nucleus of NGC 4395, which was used for tip/tilt correction during the observations,  being less bright. For the dynamical modeling, we decompose the profile into a multi-Gaussian expansion \citep[MGE,][]{EmsMonBac94}, since it allows for an easy deprojection.

\section{Structural parameters of the NSC}
\label{sec:structural_parameters}

The motions of the molecular gas in NGC 4395's nucleus are determined by the gravitational potential, which depends on the mass of the MBH and the mass distribution of the stars around the MBH. There is no evidence that dense molecular gas in the center of NGC 4395 contributes much to the mass \citep{BokSchLis11}. In order to derive the stellar mass distribution around the central black hole, we model the surface brightness of the NSC in the WFC3 images with a parametric model. For this, we use the commonly used S\'ersic model \citep{Ser68}, which is given by
\begin{eqnarray*}
I(R) = I_e \exp\left\{{-b_n\left[\left(R/R_e\right)^{\frac{1}{n}}-1\right]}\right\}.\label{eq:sersic}
\end{eqnarray*}
$R_e$ is the effective radius and $I_e$ is the surface brightness at that radius. The cuspiness of the profile is described by the S\'ersic index $n$. We will use the S\'ersic models of the NSC in the different filter bands to infer the stellar populations of the NSC (Sec.~\ref{sec:ssp_modeling}), and to generate the stellar potential (Sec.~\ref{sec:gas_modeling}). To fit PSF convolved S\'ersic models to our data, we make use of \textsc{Galfit} \citep{PenHoImp02}. 

\begin{figure*}
\begin{center}
 \includegraphics[width=0.9\textwidth,clip=]{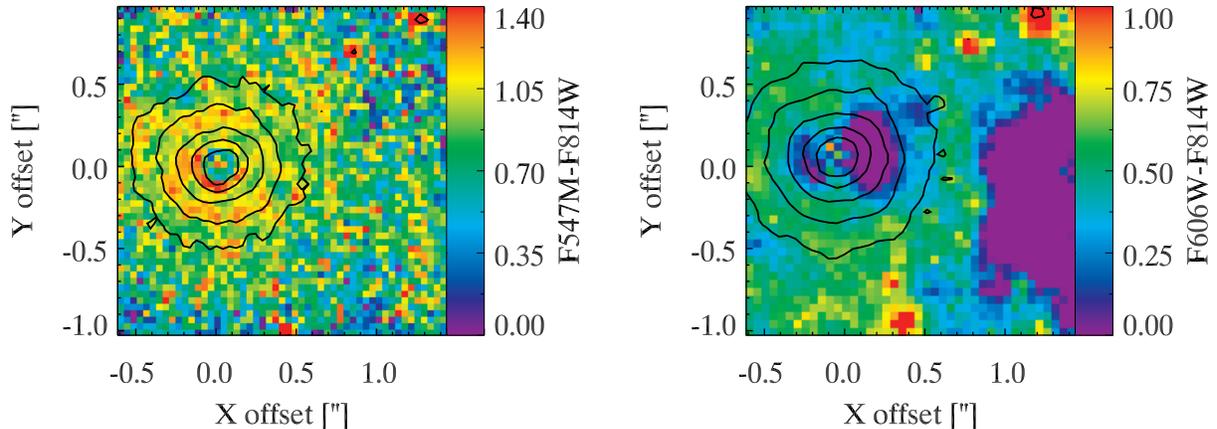}
\caption{Color maps of the nucleus of NGC 4395. On the left, we show the F547M--F814W color map based on HST/WFC3 data. On the right, we show the F606W--F814W color map based on HST/ACS data. The asymmetric blue emission is probably due to line emission.}\label{fig:colmap}
\end{center}
\end{figure*}

The modeling of the NSC is complicated by several factors. First, part of the observed emission consists of continuum emission from the central accreting black hole. Because of the small spatial scale from which this emission originates, we model this as a point source; the relative flux in each band is however not well known. Second, a blue heart-shaped feature is clearly seen extending horizontally across the NSC in the F606W--F814W color map made from the ACS data (Fig.~\ref{fig:colmap}).  The similar F547M--F814W WFC3 data shows a much less dramatic color difference and, since the emission of this double-lobed structure is similar in morphology to the line emission in the NIFS data, we thus assume that this region is contaminated by narrow line region emission from the AGN. Since the F606W filter contains [\ion{O}{3}], H$\alpha$ and H$\beta$, this emission is more prominent in the F606W--F814W color map than in the F547M--F814W colormap. Although the emission line regions are brightest in our UV filters, they are clearly seen in the residual images of the other filters. There is thus no filter free of this emission. Third, the NSC is quite compact ($R_{\textrm{eff}} \sim 4$ pc; \citealt{CarBarSet15}), making distinguishing the AGN emission from that of the NSC somewhat degenerate. Of these complicating factors, the emission line region is the most difficult to deal with; in the following subsections we use three separate methods to try eliminating this component from our fits; comparison of these methods gives us some sense of our systematic uncertainties in the fits.  We will then use these models in our black hole mass determinations in Section \ref{sec:gas_modeling}. 

For comparison, we also provide the structural parameters of the NSC obtained by \citet{CarBarSet15} from the same WFC3 data, but without dealing with the emission line region in a special way, in Table~\ref{tab:structpars}. The parameters of the NSC as determined by \citet{CarBarSet15} are sufficient for analyzing the global shape of the NSC. However, since the black hole mass is only a fraction of the NSC mass, a slight alteration of the density profile of the NSC by the emission line region could change the amount of mass at the center of the cluster, and thus change our black hole mass determination. A very accurate treatment of the emission line region is thus crucial for our dynamical modeling.

\subsection{Masking emission-line regions}\label{sec:masking}
The most obvious way to get rid of the emission line regions is by masking them out. We identify all central pixels in the F606W--F814W color map that are bluer than 0.38 mag, and create a bad pixel map for \textsc{Galfit} based on these pixels.

We fit the light distribution of the NSC with a combination of a S\'ersic profile and a point source. We first determine the S\'ersic index and effective radius in the F547M band, and then keep those fixed between different bands. We find a S\'ersic index $n=2.25$ and $R_{\textrm{eff}} \simeq 3.6$ pc. We note that these values for the S\'ersic are not atypical for NSCs: the Milky Way NSC has $n \approx 3$ and M32 has $n \approx 2.3$ \citep{GraSpi09}. 

\subsection{Iterative residual subtraction}
Instead of masking the regions, we can also try to model and subtract the emission line contribution in each filter. We tried two ways to do this. The first method, which we dub ``iterative residual subtraction'', makes use of the fact that the emission line regions show up as strongly positive residuals after fitting. For each band, we identified strongly positive residuals from our \textsc{Galfit} fit, with the bluest pixels still masked, as emission line regions. The thus created ``emission line map'' was then subtracted from the original image, which was then fit again with \textsc{Galfit}. This procedure was iterated a few times. It is a priori not clear that this procedure only identifies the non-axisymmetric emission, since fitting with the wrong NSC model can lead to the removal of stellar emission too. We were therefore very careful in identifying at which intensity to clip, which unfortunately leads to some arbitrariness.

The final structural parameter data show a somewhat more compact NSC, with S\'ersic index $n=4.4$ and effective radius $R_{\textrm{eff}}=3$ pc. The NSC is 0.07 mag brighter in the F814W band than with the masking method. Although it may seem counter-intuitive that subtraction of the residuals would lead to a brighter NSC, the overall curvature and compactness of the surface brightness profile are significantly changed, resulting in this change in brightness.

\subsection{Subtraction of the emission-line regions}
Inspection of the F275W image suggests that the narrow line emission contamination is most prominent in the F275W band, possibly due to the lack of stellar emission in these bands. Its isophotes are the most non-circular, and resemble those of the \brg\ emission in the NIR.
 Instead of masking the emission line regions, we therefore attempt to remove the emission line regions by subtracting the F275W image from the redder bands after appropriate scaling. Since the diffraction pattern in the F275W band is different from the other bands, we first subtract the central point source from the F275W band. 

To scale the emission of the F275W band to the other bands, we assume that the emission line region is the narrow line emission of the AGN. We therefore use a typical narrow line emission spectrum, from the well studied nearby Seyfert 2 galaxy NGC 1068 \citep{SpiStoBra06}, to determine the flux ratio between the F275W band and the redder bands. Subtracting the scaled point source subtracted F275W image from the other bands gave surprisingly good results in the F547M band whereas the F814W band required another 5\% scaling in the flux of the F275W band. 

The best fitting S\'ersic profile, determined in the F547M band, has S\'ersic index $n=1.4$ and $R_{\textrm{eff}} = 4.4$ pc. The luminosity of the NSC is lower than for the other two methods: compared to the masking method the NSC is 30\% fainter in the reddest band and in the bluer bands even more. We note that it is possible that some of the subtracted F275W emission is in fact stellar.

\subsection{Preferred structural model}
We summarize the structural parameters found for the four different colors using the three different methods, together with the results of \citet{CarBarSet15}, in Table \ref{tab:structpars}. 
To assess which method describes the cluster profile the best, we compare how the models fit the outer, unmasked portion of the NSC.  We avoid consideration of the inner parts of the fit, because the iterative residual subtraction gives a low $\chi^2$ by construction. We therefore decided to calculate a $\chi_{\nu}^2$, with
\begin{eqnarray}
\chi^2 = \sum_{i \in \mathrm{Good\ pixels}} \frac{\left(\mathrm{I}_i-\mathrm{Model}_i\right)^2}{\sigma_i^2} 
\end{eqnarray}
where the good pixels are all unmasked pixels (determined from the bad pixel map of the masking method) in an annulus of 4--20 pixels from the cluster center, corresponding to roughly 1--5 $R_{\mathrm{eff}}$. Here $\mathrm{I}_i$, Model$_i$ and $\sigma_i$ denote the observed intensity, model intensity and observational uncertainty of each pixel value.

\begin{figure}
\begin{center}
\includegraphics[width=0.45\textwidth,clip=]{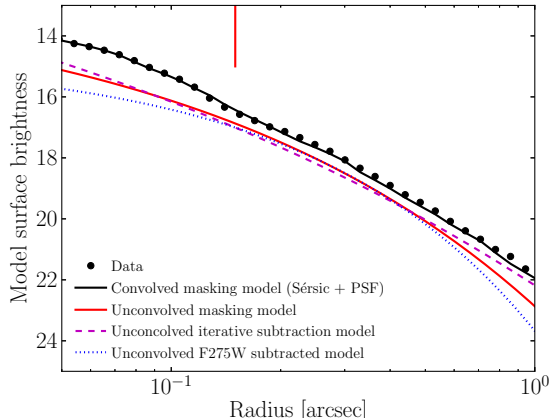}
\caption{ Unconvolved S\'ersic models of the nuclear star cluster, derived from fitting the F547M band by masking the line emission, iteratively subtracting residuals and subtracting the scaled F275W image. We also show the data overplotted on the PSF-convolved S\'ersic + point source model that was derived by masking the line emission. The red vertical line shows the effective radius of this S\'ersic model of the NSC.}\label{fig:struct_res}
\end{center}
\end{figure}

The masking and iterative residual subtraction method have the lowest $\chi^2$ per pixel values (4.3 and 4.1), the $\chi^2$ of the F275W subtraction method is slightly higher (4.7). Given the small difference in performance between the masking method and the iterative subtraction method, we adopt the simpler approach and use the results from the masking method as our preferred NSC parameters in our dynamical modeling. To get a sense of the systematic uncertainty in our BH mass measurements due to the poorly constrained NSC profile, we also perform dynamical modeling using NSC parameters from the other two methods.
The profiles of all uncolvolved S\'ersic models of the NSC are shown in Fig. \ref{fig:struct_res}, as well as a PSF-convolved model including the central point source emission.

\begin{deluxetable*}{ccccc}
\tablecaption{Best fit parameters \label{tab:structpars}}
\tablehead{\colhead{Band} & \colhead{Magnitude$^1$} & \colhead{Effective radius} &  \colhead{S\'ersic index} & \colhead{Point source magnitude}\\ 
   &   [mag]     &  [$\arcsec$]               &            &   [mag]}
\startdata
Carson et al. 2015\\
F336W & 17.61 & 0.130 & 1.86 & 17.25\\
F438W & 18.44 & 0.168 & 1.83 & 18.22 \\
F547M & 17.72 & 0.176 & 2.80 & 18.15 \\
F814W & 16.98 & 0.221 & 1.41 & 17.03 \\
\hline
Emission line masking\\
F336W & 18.42 & 0.150 & 2.25 & 16.80\\
F438W & 18.63 & 0.150 & 2.25 & 17.93 \\
F547M & 17.90 & 0.150 & 2.25 & 17.99 \\
F814W & 16.89 & 0.150 & 2.25 & 17.10 \\
\hline
Iterative residual subtraction\\
F336W & 18.17 & 0.123 & 4.43 & 17.01\\
F438W & 18.36 & 0.123 & 4.43 & 18.23 \\
F547M & 17.69 & 0.123 & 4.43 & 18.36 \\
F814W & 16.82 & 0.123 & 4.43 & 17.32 \\
\hline
F275W subtraction \\
F336W & 19.75 & 0.183 & 1.41 & 17.55 \\
F438W & 19.32 & 0.183 & 1.41 & 18.78 \\
F547M & 18.19 & 0.183 & 1.41 & 18.66 \\
F814W & 17.23 & 0.183 & 1.41 & 18.38 \\
\enddata
\tablecomments{ (1) Magnitude of the S\'ersic component in the Vega system }
\end{deluxetable*}

\section{Stellar population synthesis modeling}
\label{sec:ssp_modeling}

An important constraint on the mass of the nuclear star cluster comes from the analysis of its stellar populations. Although the magnitudes of individual bands vary depending on the fitting method and masking, some of the colors are surprisingly constant. In particular, the F547M--F814W color is 1.0, 0.87, and 0.96 for the three methods that do deal with the emission line regions. The F336W--F438W colors show however a bigger spread with F336W--F438W $= -0.21$, $-0.19$, and 0.33.

To constrain the mass of the NSC, we need to infer its stellar mass-to-light ratio (M/L) in a certain band. A common choice for this is the reddest band, since it is least sensitive to the presence of young stellar populations and extinction. We therefore infer the M/L ratio of the cluster in the F814W band.

We make use of the Flexible Stellar Population Synthesis code \citep[FSPS,][]{ConGunWhi09} to calculate colors and masses of stellar populations with a range of metallicities and ages. We select four metallicities, which cover a very metal poor scenario ($\log{Z/Z_\odot}=-1.50$), a metal poor scenario ($\log{Z/Z_\odot}=-0.69$), solar metallicity ($\log{Z/Z_\odot}=0.0$) and super solar metallicity ($\log{Z/Z_\odot}=+0.20$), and 180 different ages, logarithmically spaced between 100 Myr and 15 Gyr.

We assume  the initial mass function (IMF) of \citet{Kro01}. FSPS takes into account that the mass of the cluster evolves over time; we assume that all stellar remnants are contained within the cluster, but do not calculate the mass in gas, since it is not clear that this gas is still present. We then let FSPS make photometric predictions for four of our five optical bands. We exclude the F275W because of the large fraction of likely non-stellar emission.

\subsection{Single stellar population (SSP) fits}
As a first attempt, we fit an SSP to the magnitudes derived with the \textsc{Galfit} fits. The photometric uncertainties on the data are small, typically 1 per cent. This uncertainty does not reflect the systematic errors such as model mismatch, or remaining line emission. The uncertainty represent therefore a lower limit on the real uncertainty. 

We use the $\chi^2$ as our figure of merit for the fits. For each age and metallicity in the library, we calculate the predicted magnitudes, given a stellar mass. A mass-to-light ratio is then calculated by dividing the F814W luminosity by the mass, which can be different by a few per cent from the theoretical M/L ratio from the stellar library, since it allows for a statistical error in the F814W band luminosity.

Fig. \ref{fig:ML_single_SSP} shows a probability distribution function of the fitted M/L$_{\mathrm{F814W}}$. The best fitting single stellar population is old and metal poor. The $\chi^2$ of individual fits is not perfect, in the sense that the reduced $\chi_\nu^2 > 1$. An old metal poor population provides a better fit than a younger metal rich population. The age and metallicity of the nuclear star cluster in this scenario would be consistent with the NSC being built up from accreted globular clusters \citep{LotTelFer01,Ant13}.

\subsection{Composite stellar population (CSP) fits}
\begin{figure}
\begin{center}
\includegraphics[width=0.49\textwidth,clip=]{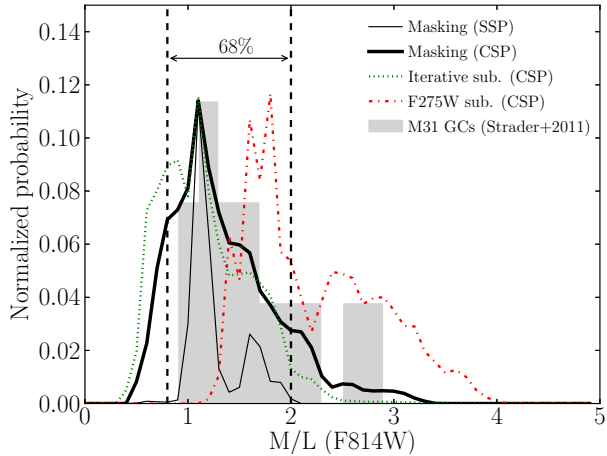}
\caption{Normalized probabilities of the mass-to-light ratios of single and composite stellar populations. An SSP fit to the colors of our favorite structural model (thin black line) prefers a relatively metal poor and old population, because old metal-rich populations underestimate the emission in the bluest bands. The range of M/L$_{\mathrm{F814W}}$  ratios for composite populations (thick line) is broader, since metal-rich populations in combination with young populations can now also fit the data. Although scaled to a common value for this plot, the likelihood of the CSP fit is higher than that of the SSP fit ($\Delta \chi^2 \approx 7$). The CSP fits to the the colors of the other two structural models are shown as a green dotted line (for the iterative residual subtraction method) and red dash-dotted line (for the F275W subtraction method). We also show as a histogram the empirical mass-to-light ratios of M31 GCs in the {\it I} band, based on data from \citet{StrCalSet11} and the Revised Bologna Catalogue }\label{fig:ML_single_SSP}
\end{center}
\end{figure}

Although it is possible that the NSC formed from old metal-poor globular clusters, sitting at the galaxy center it is very likely to have been exposed to gas accretion and subsequent star formation \citep{WalBokCha06,SetDalHod06}. A more physical mass model of the NSC should therefore include the possibility of an additional younger population. We therefore consider CSP fits where we assume that the NSC is built up from two components: an old and metal poor component, which probably contains most of the mass, and a younger component with equal or higher metallicity. For each possible combination of metallicities and ages, we calculate the best fitting weights of the two stellar populations from our four photometric bands and use this to obtain for each metallicity and age combination a likelihood and a M/L ratio in the F814W band. 
We show the normalized likelihoods of the fitted M/L$_{\mathrm{F814W}}$ ratios from composite stellar populations as a thick solid line in Fig. \ref{fig:ML_single_SSP}, together with the 68\% confidence interval. For SSP fits, M/L$_{\mathrm{F814W}}$  ratios up to 2 were allowed. For composite populations, higher M/L ratios are also allowed. The reason for this is that it is now possible to fit the red part with an old metal-rich population, which has a slightly higher mass-to-light ratio than the old metal-poor population. We note that it is possible that the emission line contamination in the blue bands mimics a young stellar population. Whether this contamination is from young stars or emission lines is in reality not very relevant though, because neither would contain much mass.

The best fit CSP models have a main component of solar metallicity and intermediate age (3 Gyr) combined with a young component. This would point at a scenario in which the nuclear stellar cluster was built up locally from enriched gas and continues forming stars, albeit at a low level.

\subsection{Empirical M/L ratios}
Observations of globular clusters suggest that SSP models with a standard IMF may not accurately predict the M/L of old systems \citep{StrSmiLar09,StrCalSet11}.  Although we do not know how the nucleus of NGC 4395 formed, the colors of the cluster are consistent with those of globular clusters, thus motivating us to examine empirical constraints on the M/L from globular clusters. 

We use the sample of GCs in M31 of \citet[][]{StrCalSet11} for which dynamical masses have been calculated. We then look up the corresponding {\it I}-band magnitudes in the Revised Bologna Catalogue \citep{GalFedBel04}, correct these for extinction, as in Strader et al., and convert the {\it I}-band magnitude to F814W magnitude. This yields 177 unique sources, for which we can determine the M/L in the F814W band. As in Strader et al., we use the virial mass as an indicator of the dynamical mass of the cluster.

 It is known that the mass-to-light ratio of globular clusters varies as a function of luminosity; low-mass globulars are sensitive to dynamical processes that can alter their M/L ratio \citep{KruMie09}. We therefore chose to use only globular clusters heavier than $10^6$ M$_\odot$ and with color $0.8 < V-I < 1.0$, since our {\it V--I} color determination of NGC 4395's cluster is the most robust and least affected by possible additional young stellar populations. We show the distribution of empirical M/L$_{\mathrm{F814W}}$  ratios inferred for these 14 clusters in Fig. \ref{fig:ML_single_SSP}.

The empirical M/L$_{\mathrm{F814W}}$  ratios of the GCs vary from 0.5-3.0 M$_\odot$/L$_\odot$ and peak around between 1-1.5 M$_\odot$/L$_\odot$, consistent with the M/Ls from the theoretical stellar populations analysis.  We note that including GCs with colors outside our chosen color range does not change the peak value or width of the histogram. As the sources shown here are the most massive GCs, it is possible that some of these contain MBHs and therefore have elevated M/L ratios \citep{MieFraBau13,SetvanMie14}. Nevertheless, we can conclude that the empirically derived {\it I} band M/L ratios of massive GCs and the M/L ratio of NGC 4395's NSC derived from stellar population synthesis modeling are in excellent agreement with each. Therefore, for our dynamical modeling we use the constraint from the CSP modeling that the M/L$_{\mathrm{F814W}}$ is  $ 1.3 \pm 0.6$  (1$\sigma$ uncertainty).

\section{Gas dynamical modeling}
\label{sec:gas_modeling}
The left panel of Fig. \ref{fig:h2velfield} shows the line-of-sight velocity of the molecular hydrogen gas. The kinematics show a clear rotational signal, with a position angle (P.A.) of $\sim-75\deg$ that is almost orthogonal to the P.A. of $\sim28\deg$ found for the radio jet \citep{WroHo06}. 

To find the best fitting potential, we adopt the commonly used assumption that the gas is on circular orbits around the black hole \citep{BerCapFun98,BarSarRix01,NeuCapReu07,SetCapNeu10}. In this case, the circular velocity of the gas is only proportional to the enclosed mass. We assume a spherically symmetric density profile, but note that the NSC appears to be slightly flattened  ($q\approx0.95$).
We characterize the light profile of the cluster using a spherical Multi-Gaussian Expanson (MGE) model \citep{Ben91}. We create the MGE using the code from \citet{Cap02} to fit the 1-D surface brightness profile of our \textsc{Galfit} models in the F814W band.  The MGE of our prefered structural model is tabulated in Table \ref{tab:mge}.  

\begin{deluxetable}{ccccc}
\tablecolumns{4} 
\tablecaption{Multi-Gaussian Expansion of the light profile of the NSC in the F814W band.}
\tablehead{Component number & \colhead{Surface luminosity} & \colhead{Dispersion} \\ 
 & \colhead{[L$_\odot$/pc$^2$]} & \colhead{[$\arcsec$]}} 
\startdata
1  &     92502  & 0.00082   \\
2  &     102553  & 0.0027   \\
3  &     94608  &  0.0076   \\
4  &     70342  &  0.019   \\
5  &     40710  &  0.042   \\
6  &     17737  &  0.086   \\
7  &     5653  &   0.16   \\
8  &     1288  &   0.28   \\
9  &     206  &   0.47   \\
10 &      23  &   0.75   \\
11 &      1.7  &    1.14   \\
12 &    0.08  &    1.71   \\
13 &  0.001   &   2.65   \\
\enddata
\label{tab:mge}
\end{deluxetable}
  We then derive the stellar density profile, $\rho(r)$, by deprojecting the MGE of the stellar surface brightness profile in the F814W band, $\Sigma(R)$, and multiplying by a dynamical M/L ratio:
\begin{eqnarray}
\Sigma(R) = \sum_{i=1}^N \frac{L_i}{2\pi\sigma_i^2} \exp\left[ \frac{-R^2}{2\sigma_i^2} \right],\\
\rho(r) = \sum_{i=1}^N \frac{(M/L) L_i}{(2\pi)^{\frac{3}{2}}\sigma_i^3} \exp\left[ \frac{-r^2}{2\sigma_i^2} \right],
\end{eqnarray}
where $R$ and $r$ are the projected and intrinsic radius, $L_i$ the luminosity of each Gaussian, $(M/L)$ the dynamical mass-to-light ratio of each Gaussian (which varies between dynamical models but is the same for all Gaussians of the MGE) and $\sigma_i$ the dispersion of the Gaussian component. 

The gas disk orbiting in this potential is built up from rings, each with its own position angle and major axis, but all concentric. We assume a single inclination for all rings (see below). Neither the kinematic observations nor the H$_2$ gas morphology provide any indication that the disk is eccentric.

We use \textsc{Kinemetry} \citep{KraCapdeZ06} to analyse our two-dimensional molecular gas velocity field. The algorithm assumes the velocity field can be described by a series of ellipses with varying maximum line-of-sight velocity, position angle (P.A.), ellipticity, and semi-major axis. Along each ellipse, the velocity is modulated by a cosine term. \textsc{Kinemetry} determines the best fitting set of concentric ellipses and provides us thus with a line of nodes (semi-major axis, PA, $q$) and the maximum velocity of each ellipse. The line of nodes is used as an ansatz in our modeling code to generate a model velocity field. During the course of this work, we found that our models performed better with a fixed axis ratio for all ellipses. We therefore use a single inclination for all rings. For each radius, we calculate the circular velocity using $v_c^2 = r d\Phi/dr$. For the MGE, this equation becomes:
\begin{align}
v_c^2 = G\frac{ M_{\mathrm{BH}} + M_{\mathrm{MGE}}(r)}{r},
\end{align}
with $M_{\mathrm{MGE}}(r)$ given by Eq. 49 in \citet{Cap08}. Given a ring's inclination and position angle, the line-of-sight velocity at each position along the ring is then calculated by multiplying the velocity with the cosine of the angle along the ellipse (the true anomaly) and the sine of the inclination. We determine the values of the line-of-sight velocity for a $10\times$ oversampled grid in $(x,y)$ coordinates by interpolating between the rings.

Since the H$_2$ kinematics are well described by single Gaussians, with dispersion values below the resolution of the spectrograph, we decided to predict the intensity weigted mean velocity $\left<I v\right>$  instead of the peak velocity of a line with a possibly complex structure. Besides a model velocity field $v_m(x,y)$, each model also has an associated broadening field with it, $\sigma_m(x,y)$, which describes the intrinsic dispersion of the H$_2$ gas. Inside the telescope, the light distribution of each frequency is convolved with the PSF. The observed data cube $Z$ can thus be described by
\begin{align}
Z(x,y,v) = \iint I(x,y) \phi(x,y,v) \mathrm{PSF}(x'-x,y'-y)\,dx'dy',
\end{align}
with $I(x,y)$ the (deconvolved) intensity field of the source and $\phi(x,y,v)$ a position dependent function,
\begin{align}
\phi(x,y,v) = \frac{1}{\sqrt{2\pi}\sigma_m(x,y)} \exp\left[ -\frac{(v-v_m(x,y))^2}{2\sigma_m^2(x,y)}     \right],
\end{align}
describing the velocity $v$ of the disk at location ($x, y$) and the thermal and turbulent broadening of the velocity field $\sigma_m$.
The mean velocity can thus be found by integrating $Z(x,y,v) \cdot v$. Since $v$ is not dependent on the position, we can change the order of integration:
\begin{align}
\left<I v\right> & = & \int Z(x,y,v)\,v\,dv &\nonumber \\
  & = & \iint I(x,y) v_m(x,y) &\mathrm{PSF}(x'-x,y'-y)\,dx'dy'. \label{eqn:vel}
\end{align}
 To simulate the NIFS observations, we can thus multiply the velocity field by the oversampled deconvolved intensity field of the H$_2$ gas, and convolve this with the oversampled NIFS PSF. We rebin the convolved field to the size of NIFS pixels and divide it by the convolved intensity field to find the predicted velocity for each NIFS pixel.

We note that this is computationally a significantly faster way to calculate the first moment than the full line modeling, since this involves doing several PSF convolutions for all velocities. However, this faster method will break down when the velocity  at the peak of the emission line (which is approximately what we measure by fitting a Gaussian during the NIFS data analysis) is no longer close to the average velocity of the line (as defined by Eq. \ref{eqn:vel}).

\begin{figure}
\begin{center}
  \includegraphics[width=0.49\textwidth,clip=]{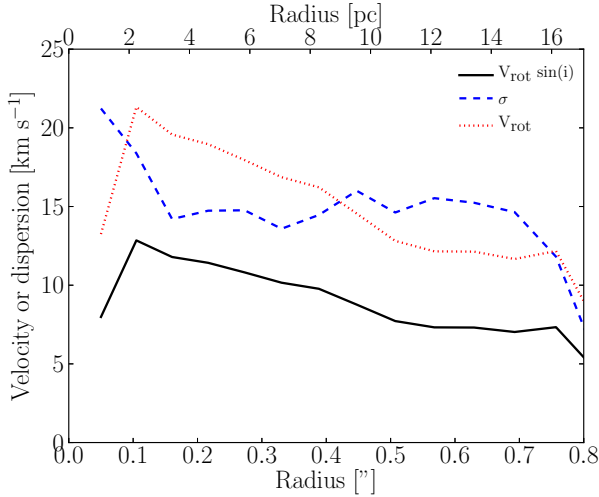}
 \caption{Maximum line-of-sight velocity ($v_{\mbox{rot}}\sin(i)$) and observed velocity dispersion ($\sigma$) as a function of radius, determined with \textsc{Kinemetry}. Values of $\sigma$ below 15 are generally not reliable, as these values are well below the instrumental resolution of NIFS ($\textrm{FWHM} = 60$ \kms). Also shown is the velocity corrected for the best fit inclination ($v_{\mbox{rot}}$).  }\label{fig:vsigma}
\end{center}
\end{figure}

If the gas disk is partly supported by internal pressure, the black hole mass in the cold disk scenario is underestimated. From Fig. \ref{fig:vsigma}, it appears that pressure could be contributing significantly. 
However, the instrumental dispersion of NIFS is $\sigma_{\mathrm{inst}}\sim25$ \kms; from previous measurements we have found that dispersion values below 15 \kms\ are not reliable. The non-central areas with the highest S/N show also the lowest dispersion values, close to 5 \kms, indicating that the dispersion in the other parts may also be lower. Full line modeling of our best fit model shows that there is no need for an additional pressure term.

\subsection{Fitting}
Our dynamical models have the following free parameters:
\begin{enumerate}
\item The dynamical M/L of the nuclear star cluster in the F814W band. We use a logarithmic scale between 0.5 and 5.
\item The mass of the IMBH. We assume that this mass is between $10^3$ and $10^7$ M$_{\odot}$ logarithmically scaled.
\item The inclination of the system. Although \textsc{Kinemetry} provides us with an ellipticity of each ring, which we convert to an inclination, we also try out configurations with a constant (possibly negative) angle added to all rings, and with all rings fixed to a single inclination value.
\end{enumerate}
Each point in this parameter space corresponds to a dynamical model. The likelihood of each data pixel is calculated from the squared difference of the observed and predicted velocity divided by the uncertainty of the observed velocity.  In Fig.~\ref{fig:h2velfield} a bright region is visible west of the nucleus, at offset 0.5--1.0 arcsec. We exlude therefore all pixels with X offset higher than 0.5 arcsec from the fit. We also exclude velocities with uncertainties higher than 5 \kms. For these uncertainties, fluctuations in the sky can cause systematic errors in our velocity determinations. This mainly affects isolated pixels of low S/N in Fig.~\ref{fig:h2velfield}.

\subsection{Results}
\begin{figure*}[t!]
\begin{center}
\includegraphics[width=1.0\textwidth,clip=]{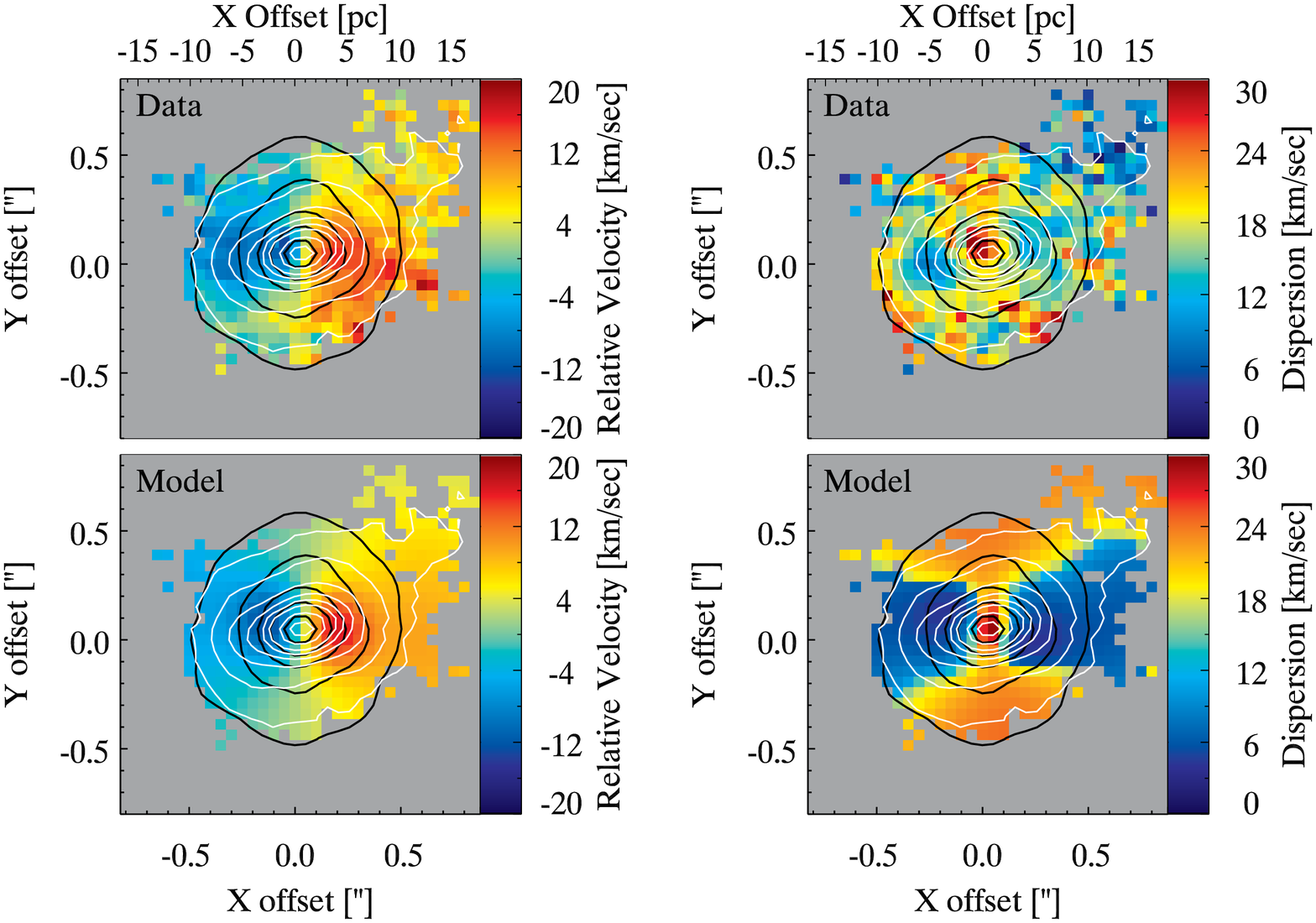}\\
\includegraphics[width=.45\textwidth,clip=]{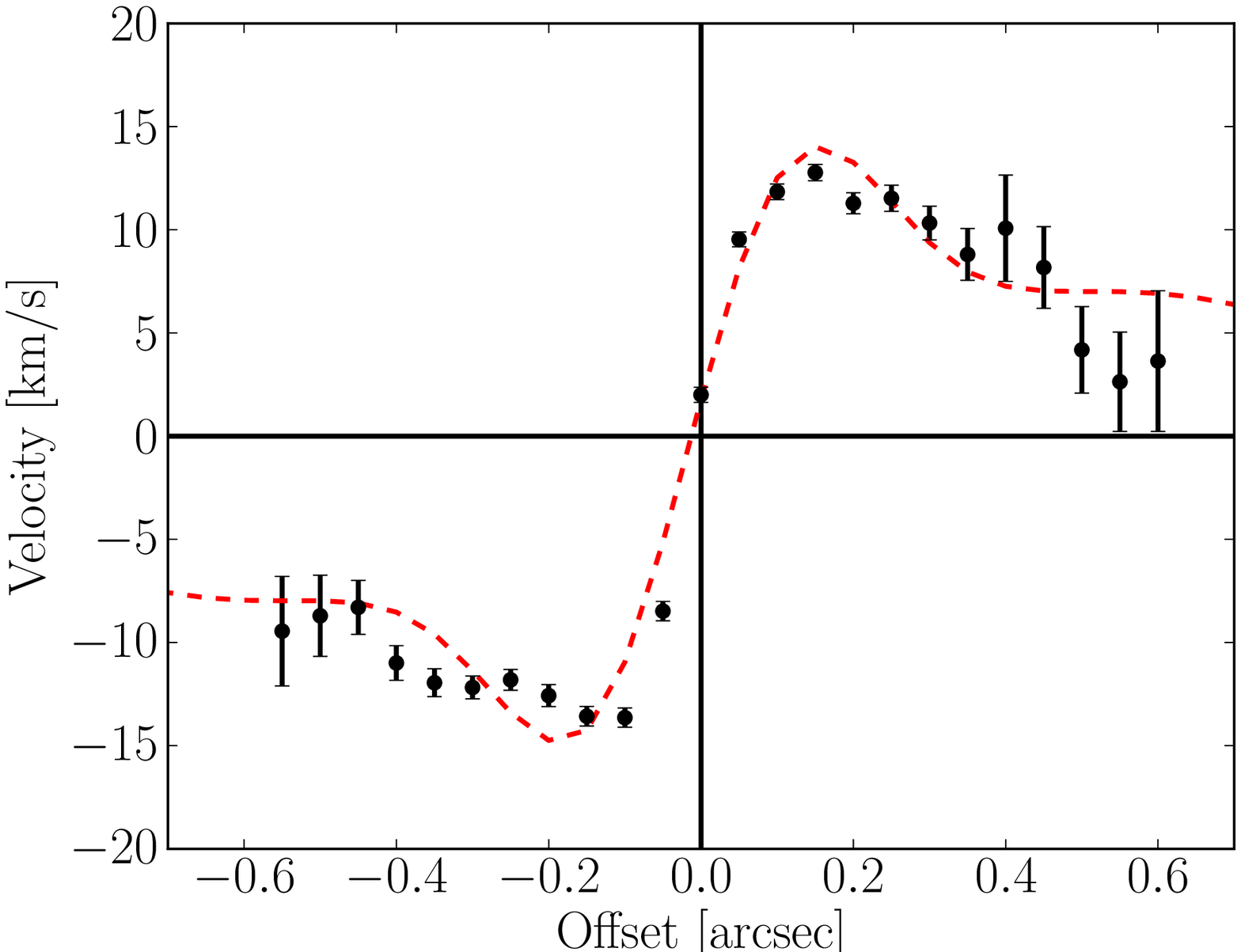}
\includegraphics[width=.45\textwidth,clip=]{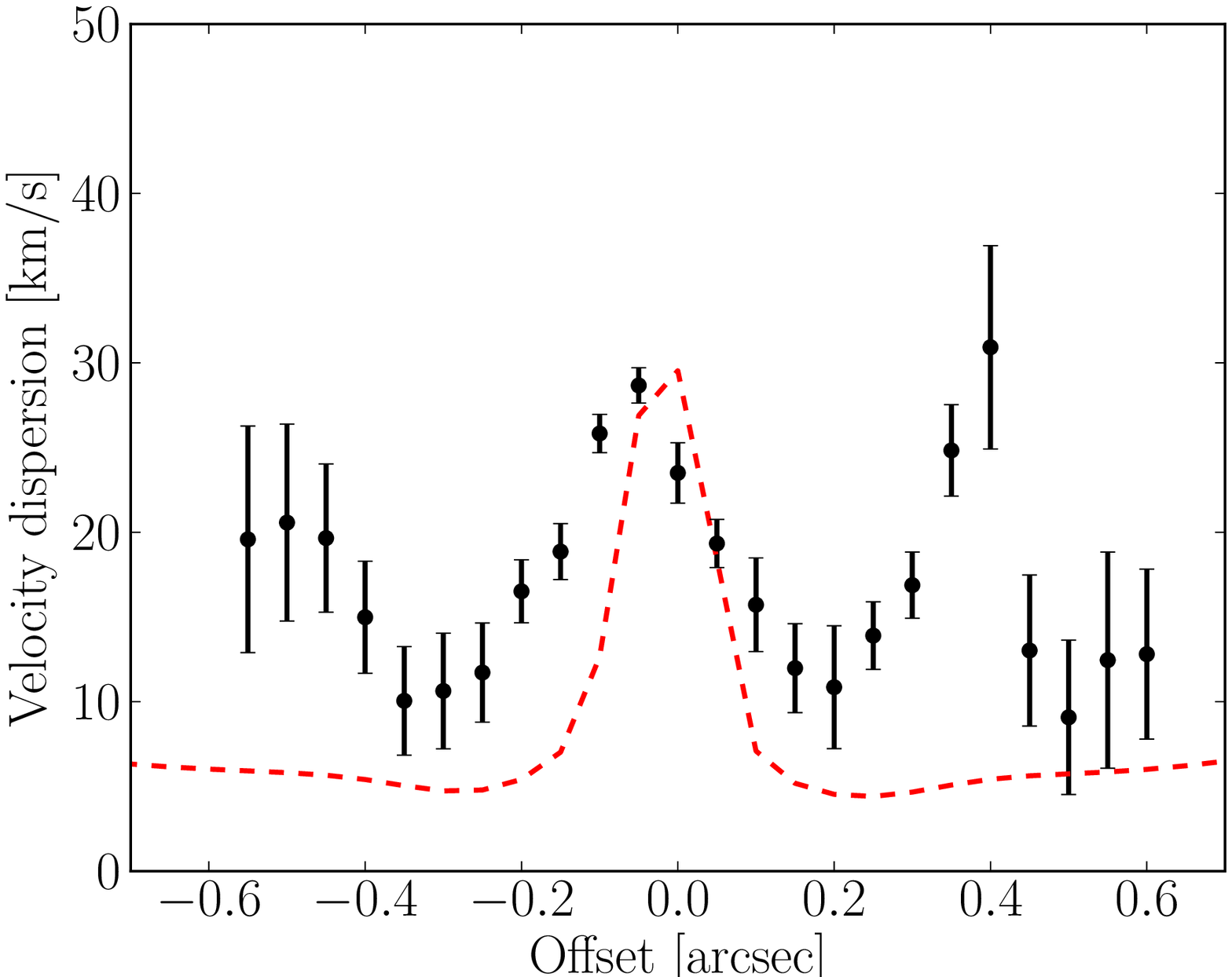}
 \caption{The same as Fig. \ref{fig:h2velfield}, but now for the model velocity and dispersion field.  In the bottom panels we show a horizontal cut through the velocity and dispersion fields. The data are shown with black error bars, and the model fields are shown with a red dashed line. Although the dispersion was not fit in our models, the peak is reproduced at the right magnitude. Note that we consider values for the velocity dispersion $\sigma < 15$ \kms to be unreliable. }\label{fig:h2model}
\end{center}
\end{figure*}
\begin{figure*}
\begin{center}
\includegraphics[width=0.49\textwidth,clip=]{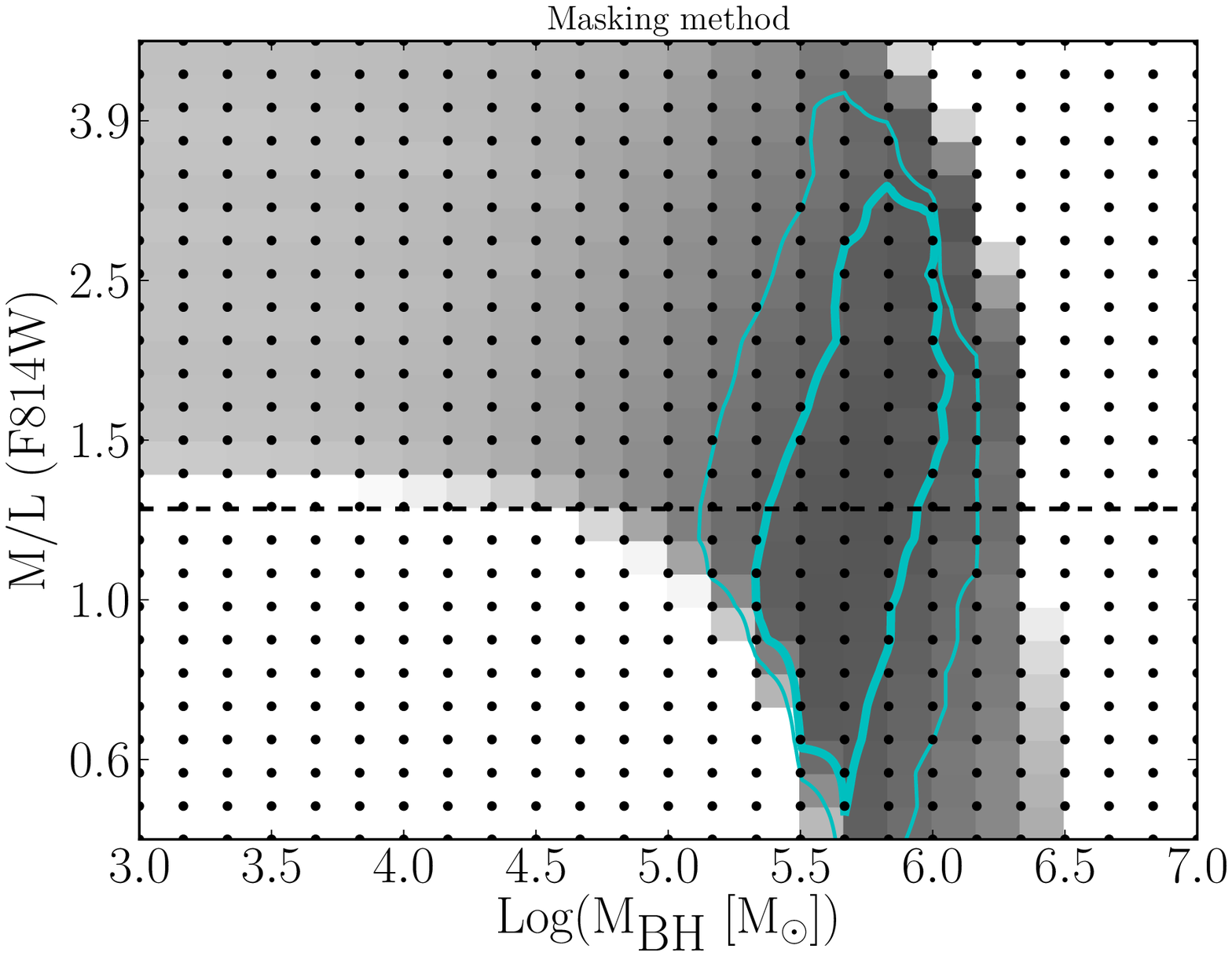}
\includegraphics[width=0.49\textwidth,clip=]{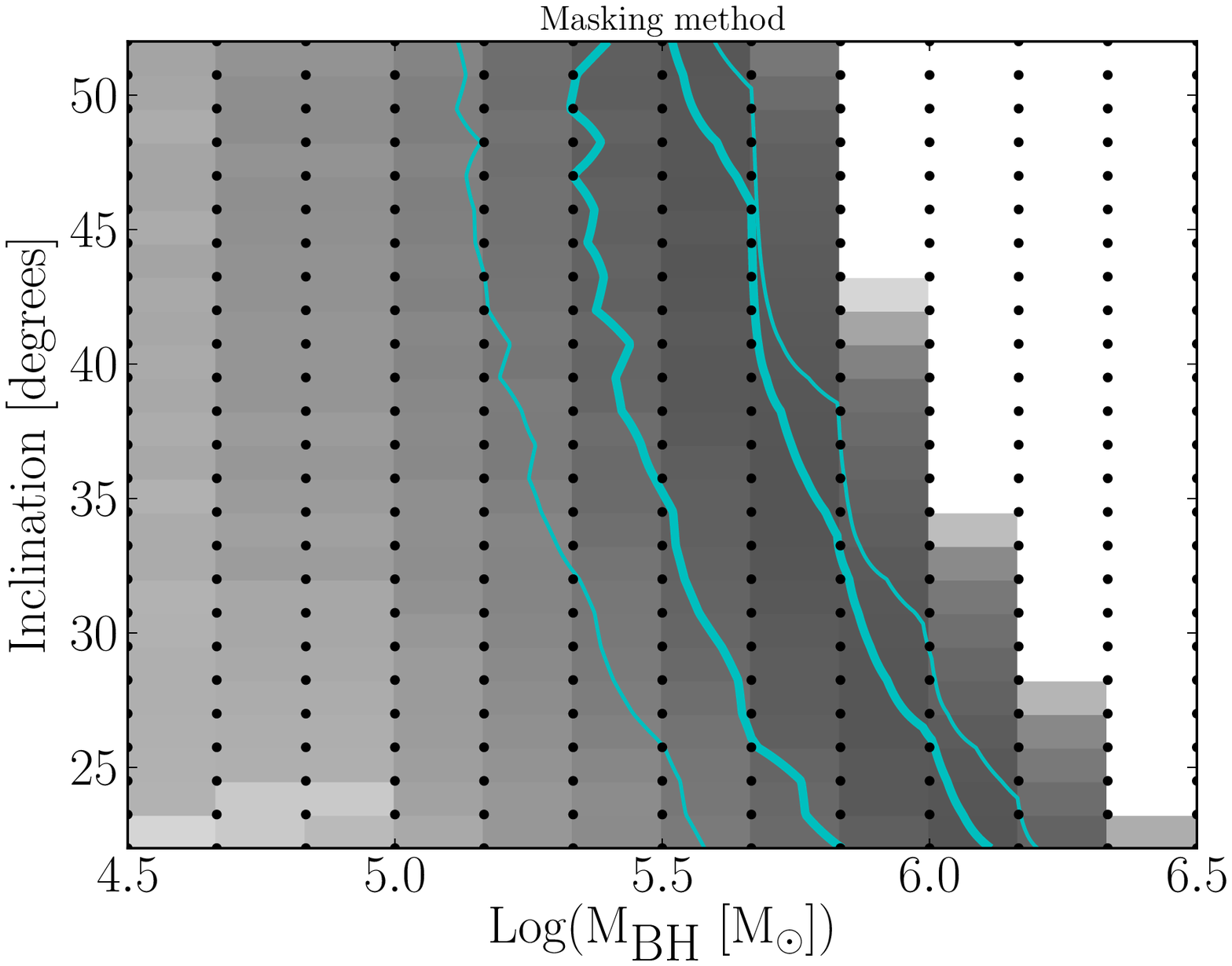}
\caption{Results of dynamical models while using photometric model B. The plots show the likelihood of the data for models with free inclination, M/L ratio and black hole mass plotted {\bf Left:} as a function of M/L ratio and black hole mass, maximized for inclination. {\bf Right:} as a function of inclination and black hole mass, maximized for M/L. Contours encompass models within 1 and 3$\sigma$ distance from the best-fit model.}\label{fig:bpmap2_dynmod}
\end{center}
\end{figure*}

We created dynamical models for 25 different inclinations, each sampling 25$\times$25 models in black hole mass and M/L. The M/L and black hole mass were sampled logarithmically. Fig. \ref{fig:bpmap2_dynmod} shows the likelihood of these models, projected by taking the maximum likelihood along the inclination axis or the M/L axis. Our dynamical models seem to prefer a configuration in which the gas disk has M/L$_{\mathrm{F814W}} = 1.3$ and inclination of 37\deg. The M/L$_{\mathrm{F814W}}$ value is in good agreement with the stellar population analysis above.  The reduced $\chi_\nu^2$ of our best fit model is 3.9. A $\chi_\nu^2 $ value $ > 1$ is not unexpected with our simplifying modeling assumptions. We therefore scale the $\chi^2$  of our models so that the best fit model has $\chi_\nu^2 = 1$. For our 3 $\sigma$ confidence limits, we select models for which the rescaled $\chi^2$ is within $\Delta \chi^2 = 14$ (3 degrees of freedom) of our best fit model's $\chi^2$.

\begin{figure}
\begin{center}
\includegraphics[width=0.49\textwidth,clip=]{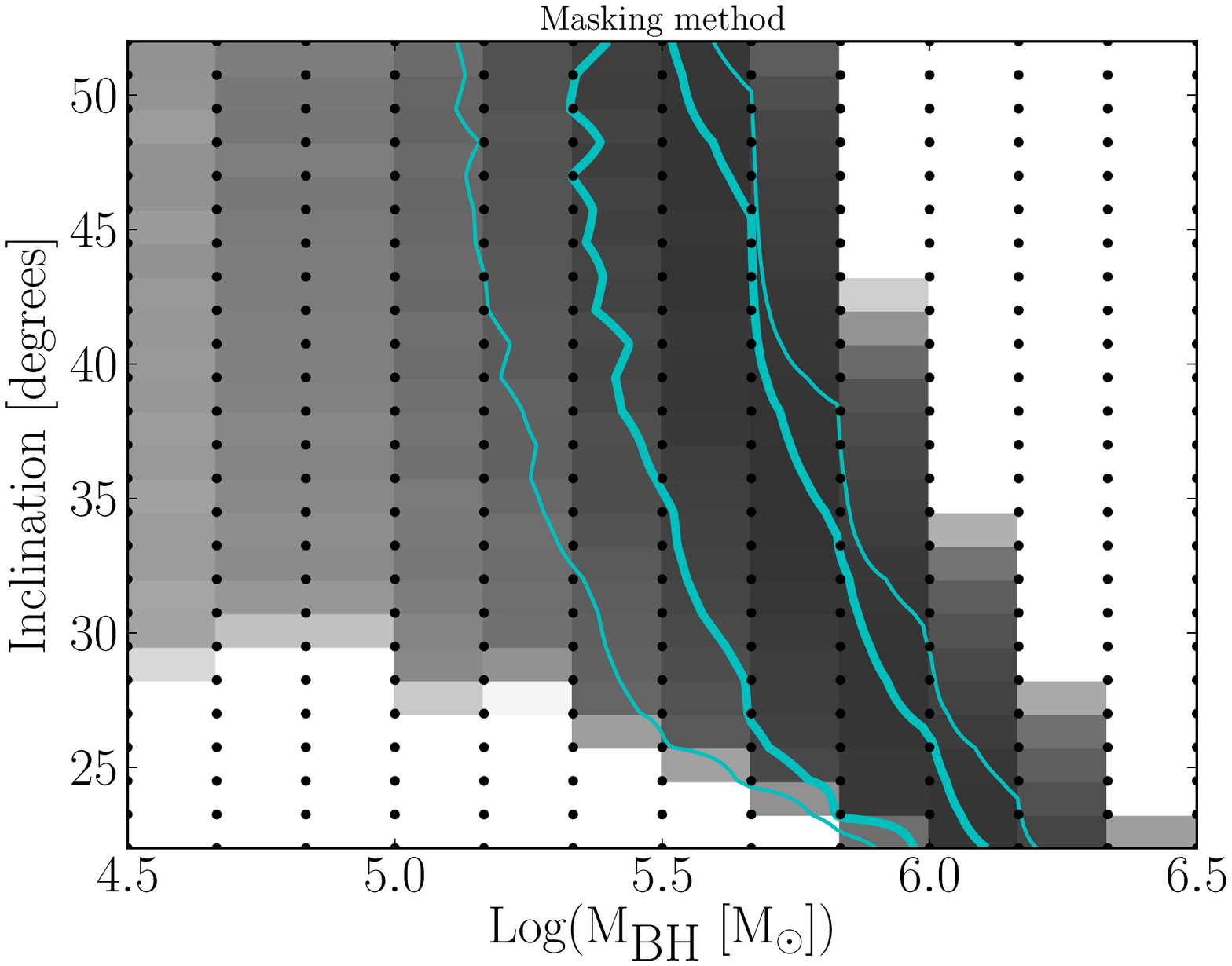}
\caption{Results of dynamical models combined with constraints from stellar population synthesis models. }\label{fig:bpmap2_dynmodml}
\end{center}
\end{figure}

We have combined the results of the composite stellar population models with dynamical constraints. In Fig. \ref{fig:bpmap2_dynmodml}, we show the results of this combination. The 1-$\sigma$ contours do not change much compared to the fits based on dynamical models only, since the best fit dynamical models agreed almost exactly with the best fit stellar population synthesis models, the joint analysis helps to exclude models with low inclination and low black hole mass. From the combined modeling of the stellar populations of the NSC and the dynamics of the molecular gas, we find a best fit black hole mass of $M=4_{-3}^{+8}\times 10^5$  \Msun (3$\sigma$ uncertainties). The velocity field of the best fit model is shown in Fig. \ref{fig:h2model}. For the parameters of this model, we also did the full line modeling, so that we could calculate the dispersion field. The central dispersion is accurately reproduced by this model, together with the depressions in dispersion along the H$_2$ disk, without adding any intrinsic dispersion for the gas. This shows that our modeling of a cold disk is justified. 

The modeling results for different structural parameters of the NSC are shown in Figs. \ref{fig:bpmap1_dynmod} and  \ref{fig:sub275_dynmod} in the Appendix. For the iterative residual subtraction method, we find a BH mass similar of  $M=3_{-2.4}^{+7}\times 10^5$  \Msun, which is very close to the mass found using the masking method. The F275W subtraction method for determining the potential gives a higher black hole mass of  $M=7_{-5}^{+8}\times 10^5$  \Msun. 

\section{Discussion and summary} 
\label{sec:discussion}
\subsection{Uncertainties and systematic errors in our mass determination.}
Several components of our modeling procedure are prone to systematic errors; the structural parameters of the cluster are uncertain, we are make simplifying assumptions about the shape of the potential, and it is possible that the observed gas does not be trace the gravitational potential. 

The largest uncertainty comes from our poor knowledge of the structural parameters of the cluster. Despite this, dynamical modeling with the different structural models from Sec. \ref{sec:structural_parameters} for the potential of the NSC yields best fit black hole masses that agree with our preferred model to within 1-$\sigma$. However, our analysis relies critically on the assumption that we can separate the continuum emission of the AGN, which is L$_{F814W} \approx 1.3\times10^6 $ L$_\odot$ from the stellar emission of the cluster L$_{F814W} \approx 1.6\times10^6 $ L$_\odot$. 

The kinematic signature observed in the H$_2$ gas does not necessarily have to be a rotating disk; it could also be tracing a weak outflow. However, there are a number of lines of evidence that suggest this is unlikely.
\begin{enumerate}
\item The H$_2$ emission in other Seyfert galaxies is known to trace gas disk components \citep[e.g.][]{RifStoDor09,HicDavMal09,HicDavMac13}.
\item The analysis of the velocity field with \textsc{Kinemetry} shows that the velocity along each ellipse follows almost a perfect cosine. An outflow would not necessarily have the same signal.
\item \citet{WroHo06} present radio continuum observations of the central parsecs of NGC 4395 and find an elongated structure, possibly an outflow, along a PA of 28 degrees, more or less orthogonal to the rotation axis of the H$_2$ disk. The orientation of this jet structure suggests  a disk interpretation for the H$_2$ gas instead of an outflow.
\end{enumerate}

In our models, we assume that the cluster is spherical and that the gas is not pressure supported. Loosening the first assumption could lower the black hole mass by a few percent. 
Introducing a pressure component will increase the black hole mass.

\subsection{Comparison to previous measurements}

With the exception of M32 \citep{VerCapCop02}, NGC 7457 \citep{GebRicTre03} and M60-UCD1 \citep{SetvanMie14}, very few dynamical mass measurements of massive black holes in galaxies with low velocity dispersions exist \citep{KorHo13,McCMa13}. For most galaxies with dispersion below 100 \kms, masses have been determined using the reverberation mapping method. The formula to determine the mass is usually written as:
\begin{eqnarray}
M_{BH} = f \frac{ (c \Delta t)(\Delta v)^2}{G}
\end{eqnarray}
where $\Delta t$ is a delay time between variation of continuum and line. The first bracketed term is thus a measure of distance from the black hole. $\Delta v$ a proxy of the average rms velocity of the line-emitting gas. For non-Gaussian line shapes, it is not clear what quantity should be used here, FWHM or $\sigma$. Virialization normalization factor $f$ is usually called the f-factor, and is often calibrated against the M-$\sigma$ relation for high-mass AGN \citep[but see][]{HoKim14}. 

Three mass  estimates for the black hole in NGC 4395 were carried out using this technique, but with different data. \citet{PetBenDes05} used the \ion{C}{4} line in the UV to estimate the time lag, and converted this to a black hole mass using $f=5.5$ (based on high-mass AGN) and using the velocity dispersion $\sigma$ of the line for $\Delta v$ (which in this case was very similar to the FWHM). This mass measurement was confirmed by \citet{DesFilKas06} using H$\alpha$ RM, although their measurement was not significant. \citet{EdrRafChe12} used broad-band reverberation mapping, which targets an individual line, in this case H$\alpha$, but uses a broad band filter to measure the variation in the line. These authors used $f=0.75$, based on a simple virial estimate of a spherical BLR, and used instead the FWHM of the line. Our black hole mass estimate is in agreement with the estimate by Peterson et al, suggesting that a high f-factor may be applicable in this particular case. 
 
In addition to the RM estimates, several additional BH mass estimates have been made for NGC 4395.  Several of these studies use estimates of the BLR size from the AGN continuum, and then use a line width to derive a mass. \citet{LirLawOBr99} estimate the black hole mass to be $5\times10^5$ \Msun (assuming a distance to NGC 4395 of 5 Mpc), based on the width of the H$\beta$ line and their estimate of the bolometric luminosity of the AGN; \citet{KraHoCre99} estimate a mass of $1.5\times10^5$ \Msun (at 2.6 Mpc distance) based on H$\beta$ and a size estimate of the BLR  from photoionization models. Using H$\beta$ and the BLR size--luminosity relation of \citet{KasSmiNet00}, \citet{FilHo03} estimate a mass  $10^4$  \Msun.

A relation between black hole mass, bolometric luminosity and break frequency of the X-ray power spectrum has also been developed \citep{McHKoeKni06}. In NGC 4395 the break frequency was measured to be $\sim2\times10^{-3}$ Hz \citep{VauIwaFab05}. If we use the bolometric luminosity from \citet{LirLawOBr99} ($1.2\times10^{41}$ erg s$^{-1}$), we find a black hole with mass $M=5\times10^4$ \Msun, which is significantly lower than our estimate. However, there are some indications that the X-ray luminosity is underestimated due to extinction \citep{NarRis11}, so that it is possible that the bolometric luminosity is in fact higher, which would increase NGC 4395's black hole mass estimate from the McHardy relation.

\subsection{A well-measured black hole mass in a bulgeless galaxy}
NGC 4395 is currently the only bulgeless galaxy with a dynamically measured black hole mass. Here we compare this galaxy to two similar bulgeless galaxies that appear to lack MBHs.

At a distance of 4.4 Mpc, NGC 4395's absolute magnitude is $M_B = -17.65$ mag and $M_V = -18.10$ mag, slightly less luminous than M33, and similar to NGC 4244, an highly inclined Sc galaxy which lies at the same distance from Earth. NGC 4244 and NGC 4395 have similar maximum rotation velocities \citep[$V_{max} \approx 90$ \kms,][]{SwaSchSan99,Oll96} which are 20 \kms\ lower than M33's. Although all three galaxies have similar morphological properties, NGC 4395 is the only galaxy with an obvious MBH. The upper limit on the mass of the MBH in NGC 4244 is $5\times10^5$ \Msun \citep{DeLHarDeb13} and for M33 is as low as a few thousand solar masses \citep{MerFerJos01,GebLauKor01}. The masses of the NSCs in M33 \citep[$1.4-2.1\times10^6$ \Msun,][]{KorMcC93,HarDebSet11} and NGC 4395 are essentially the same, whereas the NSC of NGC 4244 is 5 times heavier \citep[$1.1\times 10^7$][]{HarDebSet11,DeLHarDeb13}. Despite having a black hole at its center, NGC 4395 has succeeded in producing an ordinary NSC with a mass typical of those in late type galaxies \citep{SetAguLee08}. NGC 4395's {\it I}-band central surface brightness is, as expected, slightly lower than M33's. The only obvious difference between NGC 4395 and these other two galaxies is the kinematic lopsidedness of NGC 4395. Lopsidedness is however not an uncommon feature among late-type spiral galaxies \citep{MatvanGal98}.

 It is interesting to see how NGC 4395's black hole mass fits in with scaling relations of MBHs in more massive galaxies. NGC 4395 is not a dispersion dominated galaxy, and we do not know the value of its velocity dispersion. Comparing it to the M-$\sigma$ relation would thus not be very useful. However, \citet{KorHo13} derive a relation between black hole mass and the {\it K}-band luminosity of the bulge (M-L), with which we can compare our measurement.
In Appendix \ref{apx:decomposition} we perform a photometric decomposition of NGC 4395. Although we believe that the central upturn in the light profile of NGC 4395 is due to a bar, we can use the luminosity of the bar as an upper limit on the luminosity of the bulge, and compare the black hole mass with this upper limit. We do not know the {\it K}-band luminosity of the bar. To convert our {\it i}-band measurement of $L=4.1 \times 10^7$ L$_\odot$, to {\it K} band, we use $i - I = 0.4$ \citep{JorGreAmm06} and $I - K = 2$, which is common for old stellar populations, and is probably redder than the actual color of the bar \citep{PeldeG98}. Formula 2 in \citet{KorHo13} then predicts a black hole mass $M = 1.6 \times 10^5$ \Msun. Given the intrinsic width of the M-L relation and the uncertainties in our black hole mass measurement, we conclude that, although a factor 3 higher, our mass measurement is not inconsistent with the M-L relation.

\subsection{Summary}
We presented dynamical models of the center of the nearest Seyfert 1 galaxy NGC 4395. NGC 4395 was the first bulgeless galaxy known to host an AGN and still has the most secure detection of an intermediate-mass black hole (M $<10^6$ \Msun), although with considerable discussion about its exact mass.

 Our high-resolution HST/WFC3 data enabled us to model the morphology and M/L ratio of the compact nuclear star cluster at the center of NGC 4395. After correcting for the asymmetric contribution of emission lines to the cluster light, we find that the cluster is almost spherical (in projection). The range of M/L ratios of the cluster that we inferred from composite stellar population models is in excellent agreement with those found for globular clusters in M31. 

Combining our HST imaging data with adaptive optics assisted near-IR integral field data of the cluster, we model the dynamics of the molecular gas. Our best fitting dynamical models contain an intermediate mass black hole with mass $M=4_{-3}^{+8}\times 10^5$  \Msun (3$\sigma$ uncertainties). We characterize the influence of the uncertainty in the stellar potential on our measurement, and find that it leads to changes in the black hole mass by less than 1-$\sigma$. 
Our mass estimate is in excellent agreement with the mass estimate from reverberation mapping by \citet{PetBenDes05}, but significantly higher than the more recent estimate from \citet{EdrRafChe12} and estimates based on accretion.

\section*{Acknowledgments}
\label{sec:acknowledgments}
M.d.B. thanks Tom Maccarone for pointing out relevant references and Jenny Greene for comments on the manuscript. A.C.S. is supported by NSF CAREER grant AST-1350389. Research by A.J.B. is supported by NSF grant AST-1412693. M.C. acknowledges support from a Royal Society University Research Fellowship. V.P.D is supported by STFC Consolidated grant \#~ST/J001341/1. Support for HST program GO-12163 was provided by NASA through a grant from the Space Telescope Science Institute, which is operated by the Association of Universities for Research in Astronomy, Inc., under NASA contract NAS 5-26555.
Based on observations obtained at the Gemini Observatory, which is operated by the Association of Universities for Research in Astronomy, Inc., under a cooperative agreement with the NSF on behalf of the Gemini partnership: the National Science Foundation (United States), the National Research Council (Canada), CONICYT (Chile), the Australian Research Council (Australia), Minist\'{e}rio da Ci\^{e}ncia, Tecnologia e Inova\c{c}\~{a}o (Brazil) and Ministerio de Ciencia, Tecnolog\'{i}a e Innovaci\'{o}n Productiva (Argentina).
Some of the data presented herein were obtained at the W.M. Keck
Observatory, which is operated as a scientific partnership among the
California Institute of Technology, the University of California and
the National Aeronautics and Space Administration. The Observatory was
made possible by the generous financial support of the W.M. Keck
Foundation. The authors wish to recognize and acknowledge the very
significant cultural role and reverence that the summit of Mauna Kea
has always had within the indigenous Hawaiian community.  We are most
fortunate to have the opportunity to conduct observations from this
mountain.
\bibliographystyle{apj} 
\bibliography{paper}
\newpage
\begin{appendix}
\section{Stellar Velocity Dispersion: NIRSPEC \& NIFS data}\label{sec:nirspec}

The stellar velocity dispersion of the nuclear cluster in NGC 4395 is
a parameter of interest for dynamical modeling of the cluster providing a constraint on the total mass of the system \citep[e.g.][]{BarStrBen09,NeuWal12}.  

To date, the only direct constraint on the velocity dispersion of this
cluster is a measurement of $\sigma < 30$ \kms\ by \citet{FilHo03}
based on high-dispersion optical spectroscopy around the \ion{Ca}{2}
triplet obtained with HIRES at the Keck Observatory.  As an attempt to
measure $\sigma$ directly, we obtained near-infrared observations with
the NIRSPEC spectrograph \citep{McLBecBen98} at the Keck II telescope.

NIRSPEC observations were obtained during the first half of two
nights, 2010 March 28--29 UT, using the NIRSPEC echelle mode with a
$0\farcs72\times12\arcsec$ slit.  Weather was clear and the seeing was
excellent, with focus runs giving FWHM of $\approx0\farcs4$ on stellar
images.  On the first night we observed NGC 4395 in the NIRSPEC-7
($K$-band) filter, and during the second night the NIRSPEC-5
($H$-band) filter was used.  The galaxy nucleus was nodded along the
slit between successive exposures for sky subtraction, with individual
exposure times of 300 or 600 s, and the slit was held at a fixed
position angle of 120\deg. The total on-source exposure time was 2.0
hr in the $K$ band and 2.3 hr in the $H$ band.  We observed A0V stars
as telluric calibrators, and several K and M giant stars as velocity
templates.  Arc lamp and flat-field exposures were obtained at the
start and end of each half-night.

\begin{figure}
\begin{center}
  \includegraphics[width=0.63\textwidth,clip=]{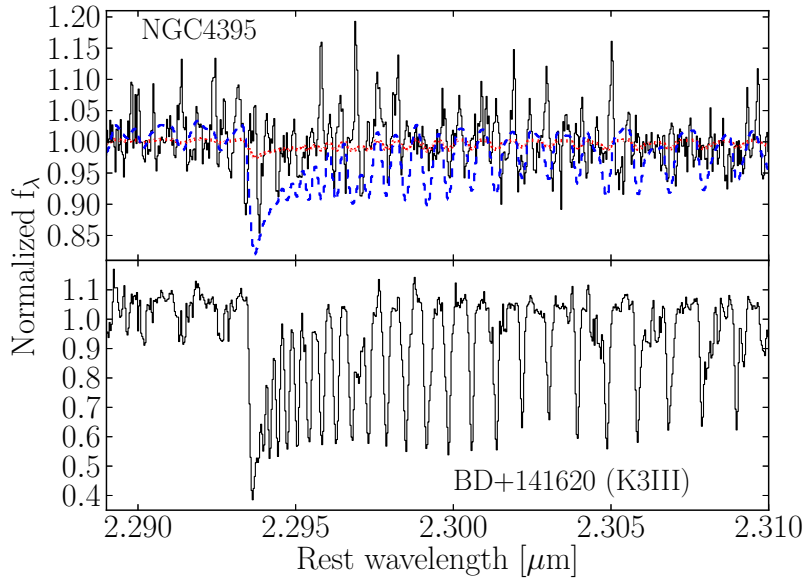}
 \caption{Normalized NIRSPEC spectra of the NGC 4395 nucleus and a K3III
star. The wavelength of the first CO bandhead is 2.2935 \micron.  We also show the expected spectrum of the nucleus based on modeling of the K3III star for 65\% (blue dashed line) and 95\% AGN contribution (red dotted line) and a velocity dispersion of $\sigma=30$ \kms. The AGN is clearly dominating the spectrum. This domination of the continuum component is also easy to discern from the Gemini/NIFS data shown in Fig. \ref{fig:nifs}, where even in the least AGN dominated aperture, the line depths of the CO bandheads are at most a few percent. }\label{fig:nirspec}
\end{center}
\end{figure}

Reductions were carried out using the NIRSPEC REDSPEC pipeline.  For
the $K$-band data, we focused on the spectral order containing the
primary CO bandhead at 2.2935 \micron\ since this is the main feature
of interest for measuring $\sigma$.  In this spectral order, the
dispersion is $3.33\times10^{-5}$ \micron\ pixel$^{-1}$.  Extracted
spectra were wavelength calibrated and flux calibrated using arc lamp
and A0V star exposures, and individual exposures were combined to give
a weighted average spectrum.  Figure \ref{fig:nirspec} shows the CO bandhead region
spectrum of the NGC 4395 nuclear cluster and of a K3III giant for
comparison.  Over this region, the NGC 4395 spectrum has S/N = 23
pixel$^{-1}$.  There is no detected CO bandhead feature in NGC 4395.  If 95\% of the spectrum were coming from the AGN, we would not be able to detect CO features at a velocity dispersion of 30 \kms. If the AGN contributes only 65\% of the flux, the features would be clearly visible.
Apparently the near-IR spectrum of the NGC 4395 nucleus is so
dominated by the AGN (and possibly by a contribution from young, hot
stars) that the CO features are diluted to the point of being
invisible in the integrated light of the cluster and AGN. This domination of the AGN continuum in the near IR is supported by the ground-based near-IR observations of \citet{MinYosKob06}, which show clear variability in NGC 4395's integrated nuclear flux by a factor $>2$.

The $H$-band spectrum is significantly noisier and also does not show any
stellar absorption features from the cluster.  
\begin{figure}
\begin{center}
  \includegraphics[width=0.63\textwidth,clip=]{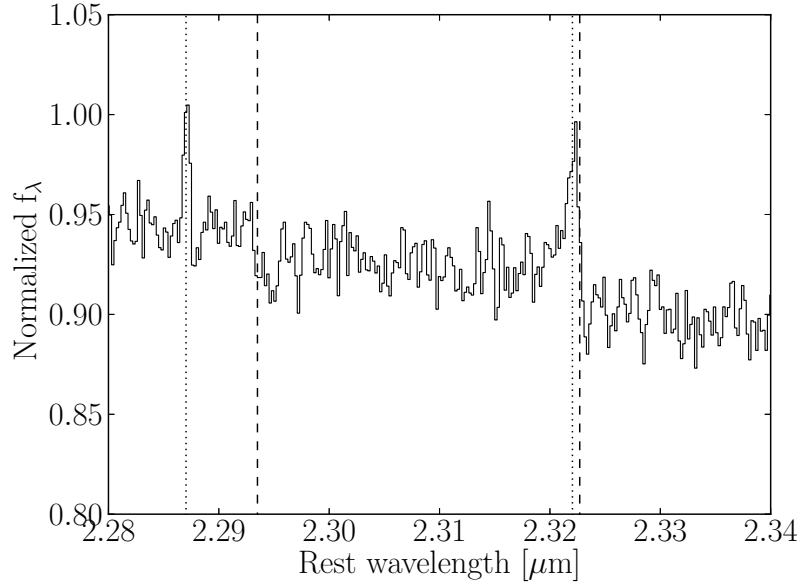}
 \caption{Normalized NIFS spectrum of the NGC 4395 nucleus. The first two CO bandheads are marked by the dashed lines. Also marked (by dotted lines) are the H$_2$ 3--2 $S(2)$ line at 2.28703 \micron\ and the \ion{Ca}{8} line at 2.32204 \micron.  }\label{fig:nifs}
\end{center}
\end{figure}

We also searched our NIFS data for signs of the CO bandhead.  We tested various annuli for signs of the CO bandheads.  Annular spectra with radii between 0$\farcs$15 and 0$\farcs$35 show the strongest signs of the first CO bandhead.  Only the first CO bandhead is seen at a depth of 3\%, and the line is at a redshift of 317~\kms, within a couple \kms\  of the redshift of the galaxy derived from \ion{H}{1} gas \citep{HayvanHog98}.  
The formal fit to the dispersion of the system suggests a dispersion of the system $<$30~\kms; however due the weakness of the line, the lack of additional CO bandhead features and \ion{Ca}{8} emission line contamination of the 2nd CO bandhead (see Fig. \ref{fig:nifs}), we regard this measurement as less reliable than the Calcium triplet measurement shown in \citet{FilHo03}.

It would still be extremely desirable to measure the velocity dispersion of this
cluster, but it is probably necessary to couple such a measurement with monitoring of the flux level of the AGN to ensure minimal dilution of the spectral features by the AGN component.
\newpage
\section{Dynamical modeling results for alternative models of the cluster}
\begin{figure*}
\begin{center}
\includegraphics[width=0.49\textwidth,clip=]{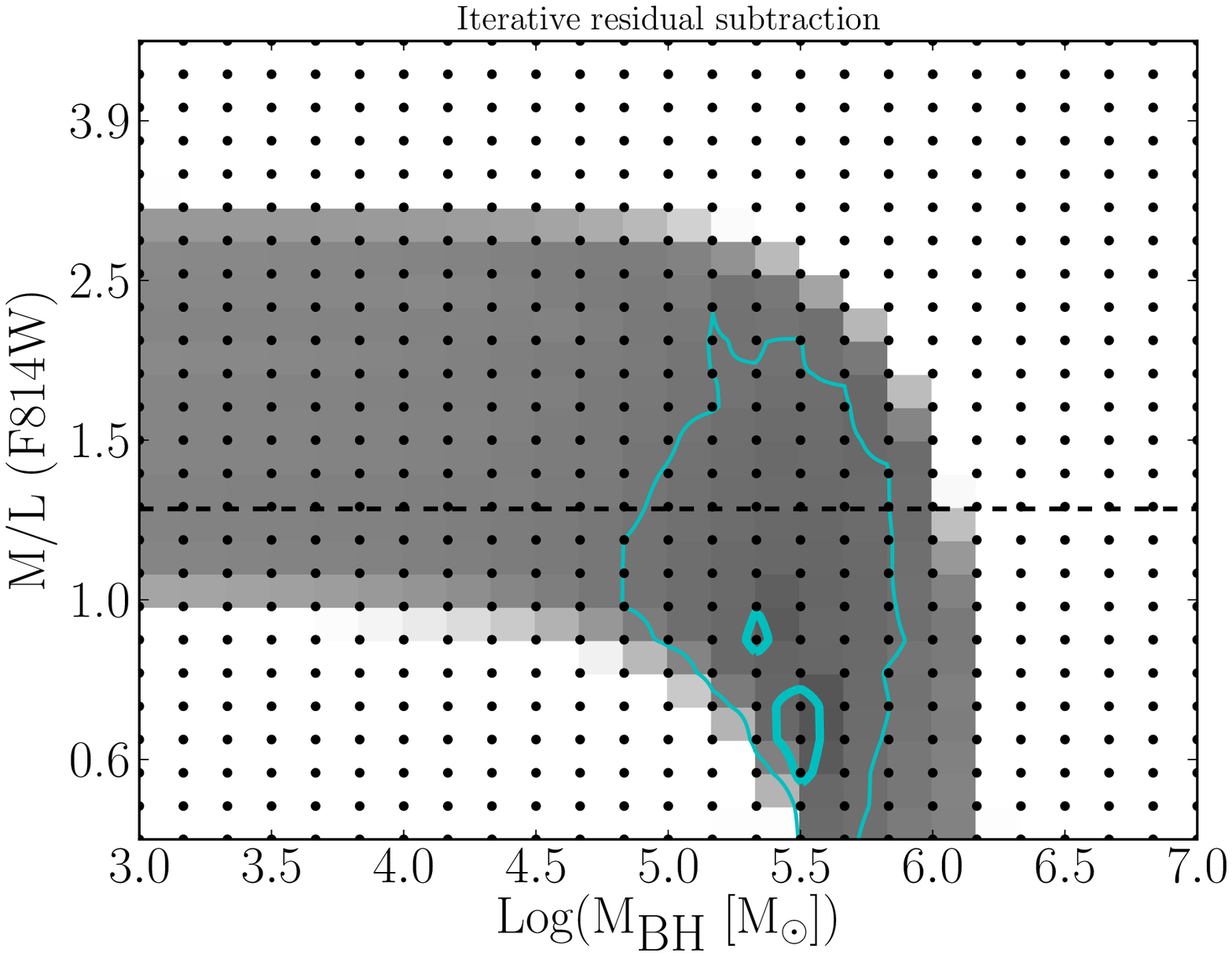}
\includegraphics[width=0.49\textwidth,clip=]{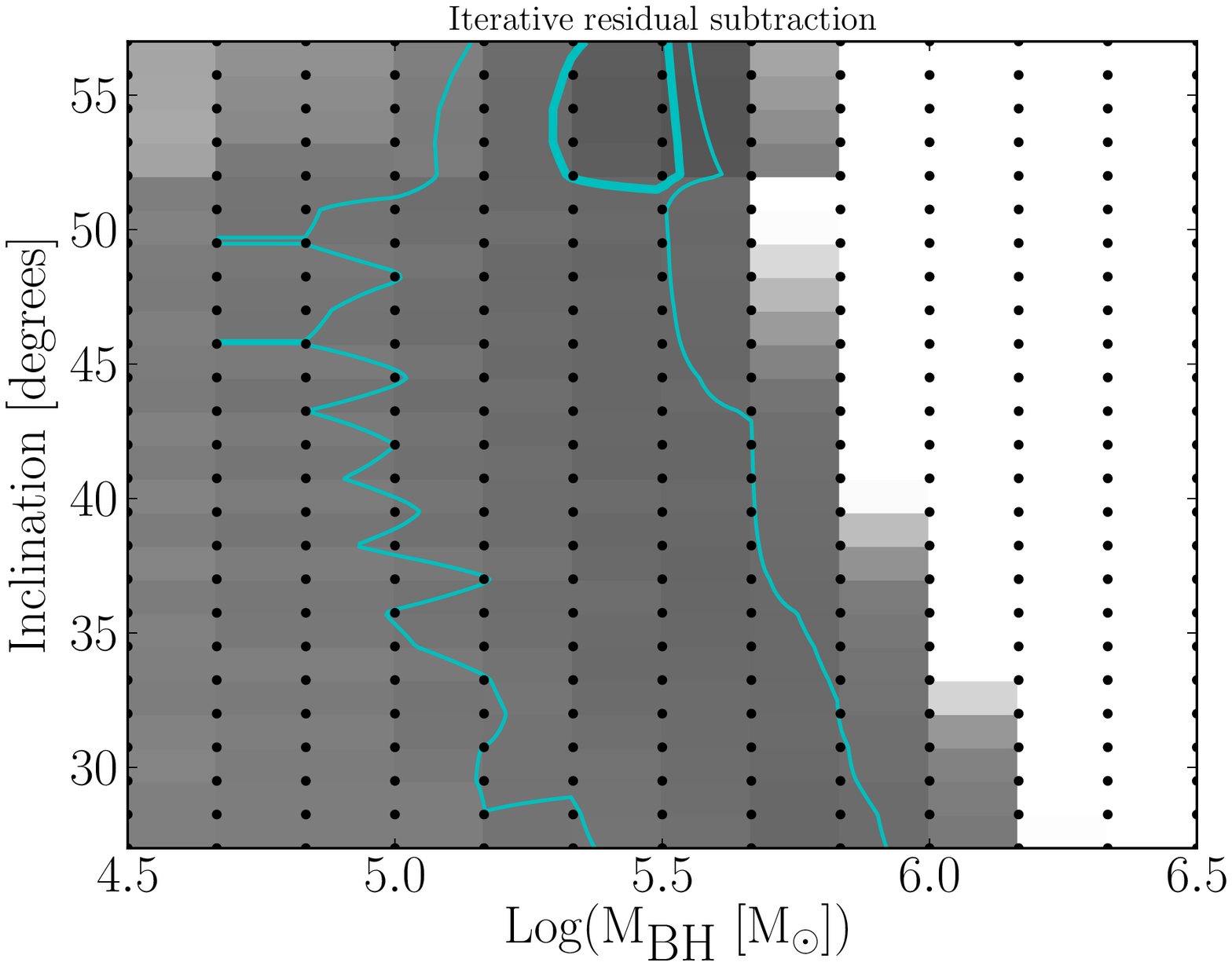}
\caption{Results of dynamical models while using photometric model B (iterative residual subtraction). The plots show the likelihood of the data for models with free inclination, M/L ratio and black hole mass plotted {\bf Left:} as a function of M/L ratio and black hole mass, maximized for inclination. {\bf Right:} as a function of inclination and black hole mass, maximized for M/L. Contours encompass models within 1 and 3$\sigma$ distance from the best-fit model. The best fit model is highly inclined.}\label{fig:bpmap1_dynmod}
\end{center}
\end{figure*}

\begin{figure*}
\begin{center}
\includegraphics[width=0.49\textwidth,clip=]{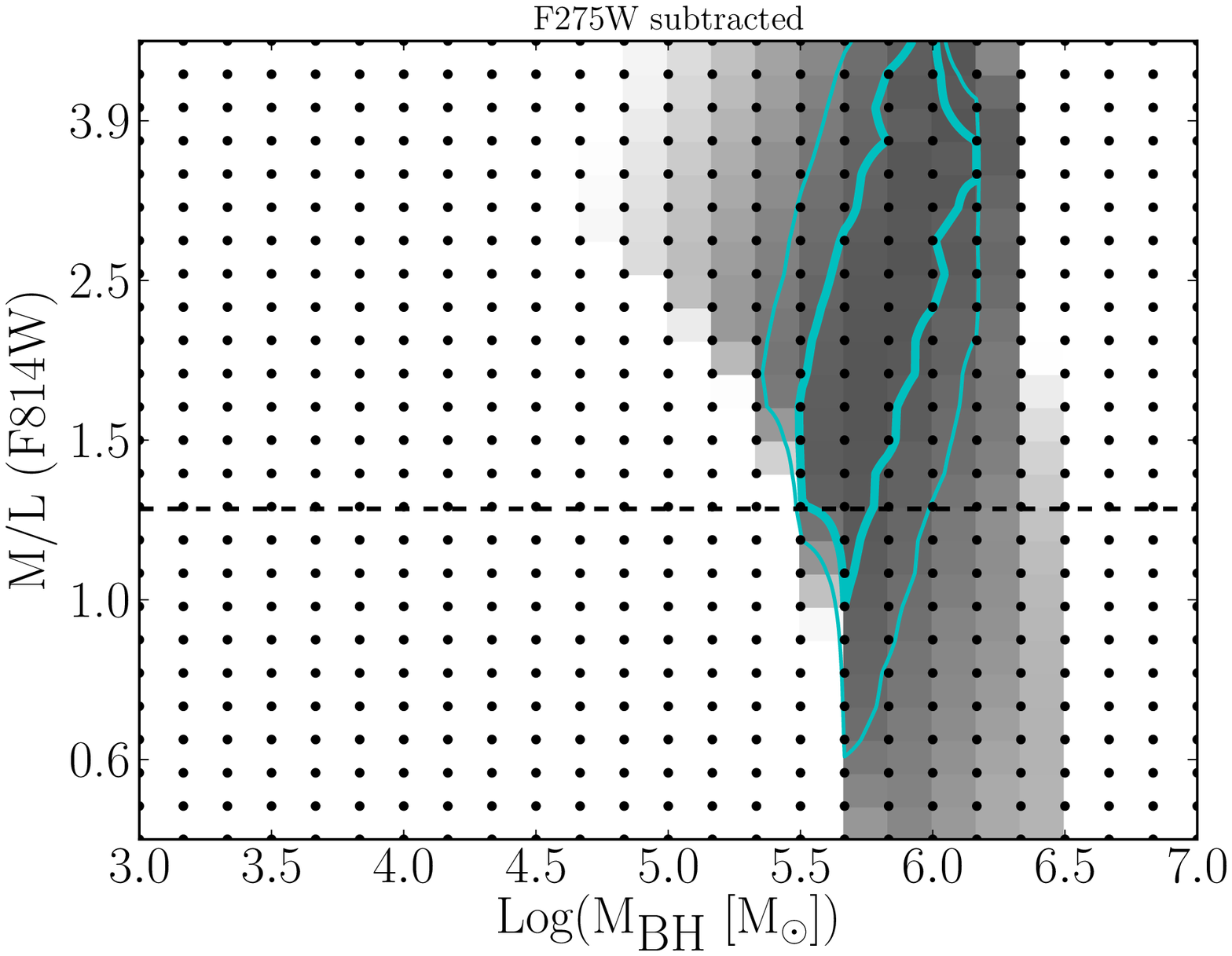}
\includegraphics[width=0.49\textwidth,clip=]{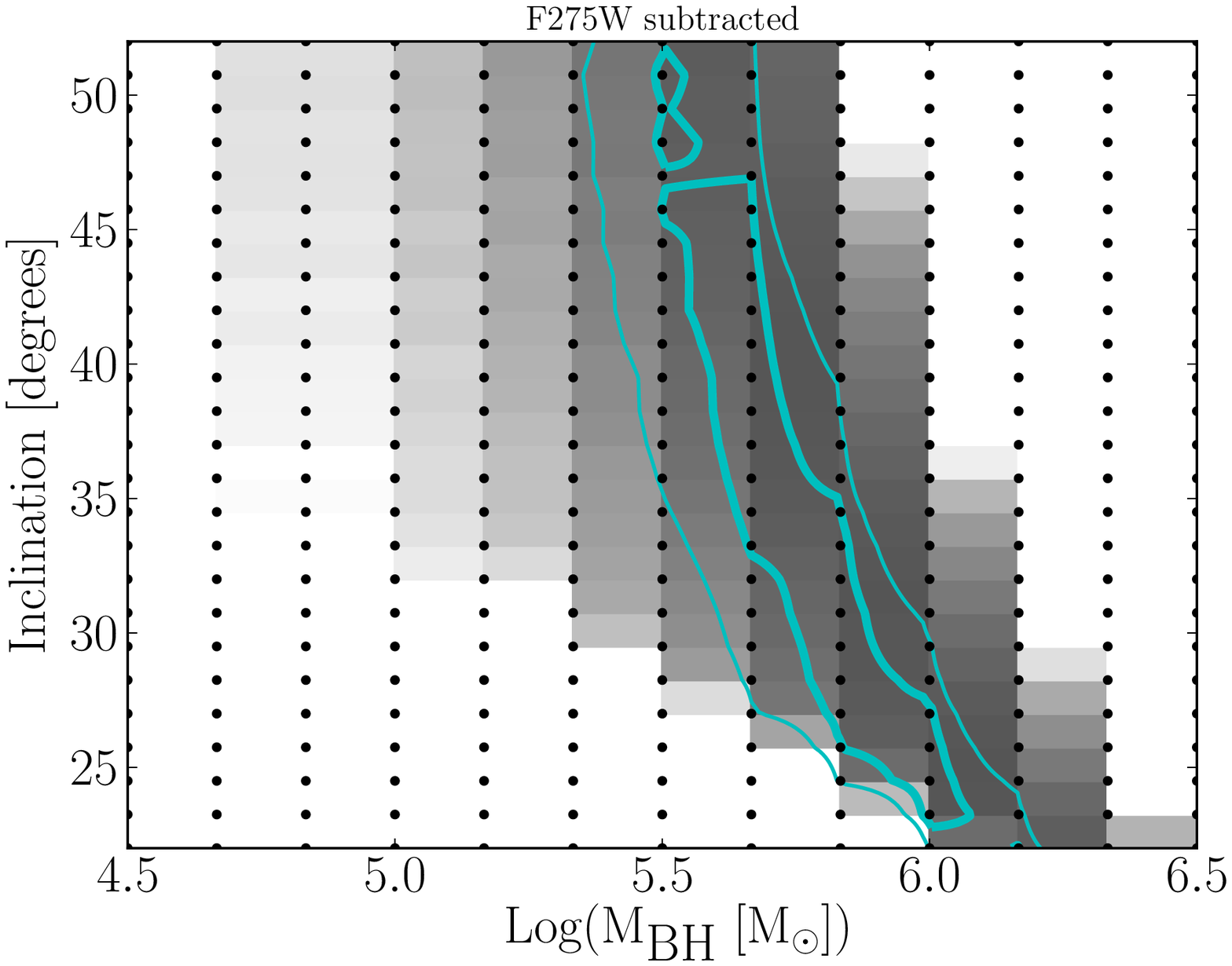}
\caption{Same as Fig. \ref{fig:bpmap1_dynmod}, but now for the star cluster model derived after subtracting the F275W image. A high black hole mass is preferred.}\label{fig:sub275_dynmod}
\end{center}
\end{figure*}

A major uncertainty in our modeling of the gas kinematics is the exact structure and mass of the stellar potential, which is difficult to measure both because of the NSC's compactness and because of the presence of line emission. In Section \ref{sec:structural_parameters}, we presented, besides ordinary masking of the emission line regions, two alternative ways to deal with them. The different estimates for the M/L ratios of the stellar populations for these alternative models are presented in Fig. \ref{fig:ML_single_SSP}. Here we present also the results of the dynamical models for these alternative models. Fig. \ref{fig:bpmap1_dynmod} shows the results of the dynamical M/L and inclination versus the MBH mass for the second structural parameter model (iterative residual subtraction). Despite the 10\% difference in NSC mass and different S\'ersic index of the profile, the black hole mass is in agreement with the one from the analysis in the main text. The modeling results using the other structural parameter model are shown in Fig. \ref{fig:sub275_dynmod}. The best-fit BH mass for this model is higher than the other two models, and simple axisymmetric Jeans models show that there is some tension with the stellar velocity dispersion measurement of \citet{FilHo03}. Possible explanations are that the F275W band contained not just line emission but also stellar emission, or that the scaling between the F275W band and the other bands is not the same for NGC 4395 and NGC 1068. 

\newpage
\section{Photometric evidence for a bulge in NGC 4395}\label{apx:decomposition}
As far as we are aware, there exists no photometric decomposition of NGC 4395 into a bulge and a disk. Such a decomposition is important to corroborate the claim that MBHs exist in bulgeless galaxies. We therefore present a photometric decomposition of this galaxy here.

We use {\it i}  band imaging data from the Sloan Digital Sky Survey  \citep[SDSS,][]{AbaAdeAgu09} to constrain the presence or absence of a bulge in NGC 4395. The data from this band are the least affected by dust and young stellar populations and less shallow than the {\it z}-band data. We note that NGC 4395 falls close to the edge of the chip.  This does not affect our analysis in the inner parts, but beyond $220\arcsec$ the data are dominated by a single irregular-looking spiral arm because of this asymmetry.

We use {\textsc{SExtractor}} \citep{BerArn96} to identify background galaxies, foreground stars and star clusters in the image and generate a bad pixel mask. We then use the IRAF\footnote{IRAF is distributed by the National Optical Astronomy Observatory, which is operated by the Association of Universities for Research in Astronomy, Inc., under cooperative agreement with the National Science Foundation.} task \textsc{ellipse} \citep{Jed87} to measure the radial surface brightness profile of the galaxy. We fit the profile with the combination of an exponential profile and a S\'ersic profile. We exclude the nucleus from the fit, as well as the data beyond $220\arcsec$. 

\begin{figure*}
\begin{center}
\includegraphics[width=0.9\textwidth,clip=]{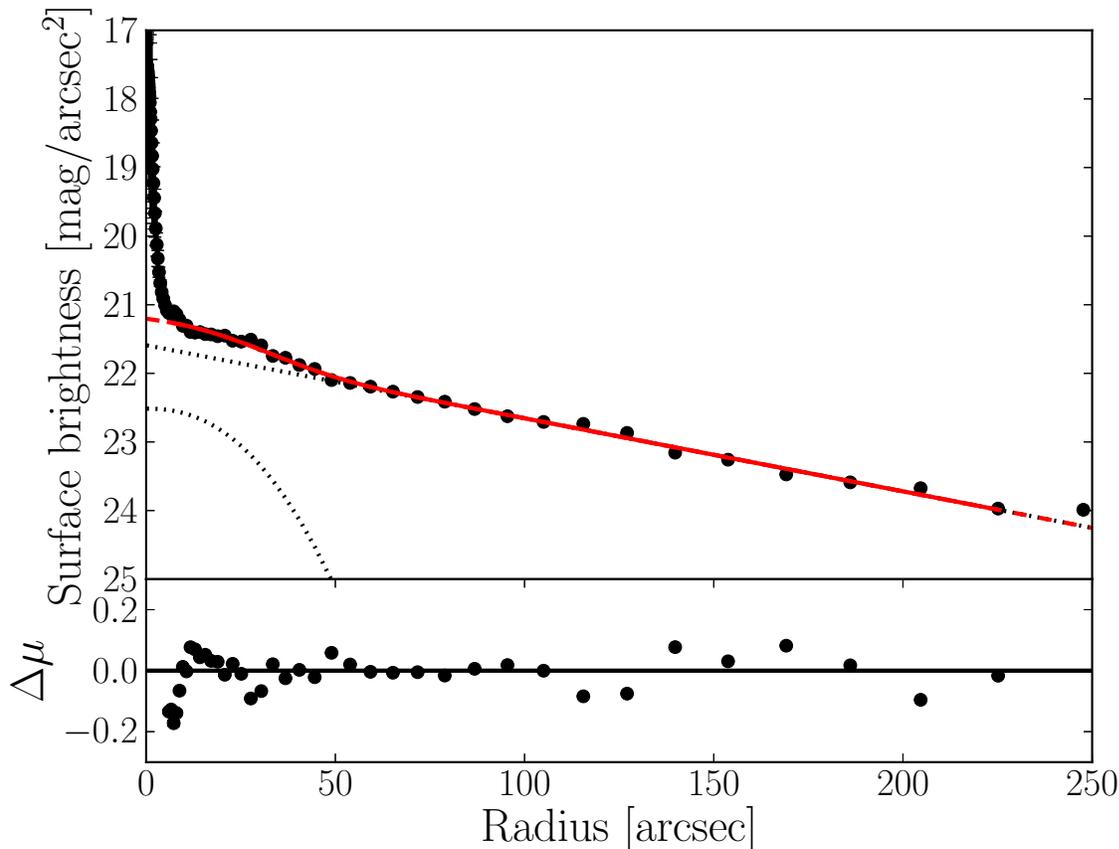}
\caption{Photometric decomposition of the SDSS {\it i} band data of NGC 4395 into an exponential and S\'ersic model. The central S\'ersic model has an effective radius of $26\arcsec$ (0.5 kpc), and a S\'ersic index $n = 0.43$.}\label{fig:photmodel}
\end{center}
\end{figure*}
Fig. \ref{fig:photmodel} shows the decomposition of NGC 4395 into a S\'ersic and an exponential model. The exponential model has magnitude m$_i = 10.3$ mag (M$_i = -17.9$ mag), scale length $h = 102\arcsec$ (2.2 kpc) and dominates over the S\'ersic profile at all radii. The S\'ersic component has  m$_i = 13.7$ mag (M$_i = -14.5$ mag), effective radius R$_{\textrm{eff}} = 26\arcsec$ (0.5 kpc) and S\'ersic index $n = 0.44$. The low S\'ersic index and low luminosity (the bulge-to-total ratio B/T $= 0.04$) indicate that this is probably not a classical bulge \citep{FisDro08}, but more likely a bar. Visually, the elongated shape suggests that it is indeed a bar, and we also find tentative support for this by a small change of 10 degrees in the position angle in the inner 40 arcsec of NGC 4395.

\end{appendix}

\end{document}